\documentclass[journal]{IEEEtran}
\usepackage{amsmath,graphicx}
\usepackage{framed,multirow}
\usepackage{makecell}
\begin{document}
\title{Medical Instrument Detection in Ultrasound: A Review}
\author{Hongxu Yang, Caifeng Shan, Alexander F. Kolen, Peter H. N. de With%
\thanks{Hongxu Yang and Peter H. N. de With are with the Department
of Electrical Engineering, Eindhoven University of Technology, Eindhoven, The Netherlands.}
\thanks{Caifeng Shan is with Shandong University of Science and Technology, Qingdao, China.}
\thanks{Alexander F. Kolen is with Philips Research, Eindhoven, The Netherlands.}}

\maketitle

\begin{abstract}
Medical instrument detection is essential for computer-assisted interventions, since it facilitates the radiologists to find the instrument efficiently with a better interpretation, thereby improving clinical outcomes. This article reviews image-based medical instrument detection methods for ultrasound-guided operations. First, we present a comprehensive review of instrument detection methodologies, which include both traditional non-data-driven methods and novel data-driven methods. The non-data-driven methods were extensively studied prior to the era of data-driven methods, such as machine learning approaches. We then discuss some important clinical applications of medical instrument detection in ultrasound, including the delivery of regional anesthesia, biopsy taking, prostate brachytherapy, and catheterization, which were all validated on clinical datasets. Finally, the principal issues and potential research directions are summarized for the computer-assisted intervention community for the future studies.
\end{abstract}


\IEEEpeerreviewmaketitle
\section{Introduction}
With increasing financial pressure on the healthcare system, there exists a general trend toward efficient workflow, shortening procedure time, and higher clinical outcome resulting in fewer repeats for any given intervention or surgery. To guide intervention operation, e.g., cardiac intervention and needle biopsy, advanced medical imaging systems such as ultrasound and fluoroscopy, are required. The imaging system offers radiologists visualization and measurement of anatomical structures. It displays the interventional activities within or outside the operation regions. The imaging system also enable the guidance of medical instruments inside the patient body without an open incision, which is commonly known as image-guided minimally invasive intervention \cite{douglas2001ultrasound,germano2002advanced,peters2006image,cleary2010image}. This approach is being increasingly adopted in many surgical applications because of its lower risk of complications, shorter patient recovery time, and therefore lower cost. Moreover, with advanced hardware and image recording techniques, it is possible to obtain operation-related data efficiently with higher quality, which enables researchers to analyze the data and provide solutions to guide the procedure. 

The medical imaging systems and the amount of medical data are rapidly growing in the past years. These developments in medical imaging enable novel applications in the area of minimally invasive surgery. Among the key imaging modalities used in image-guided minimally invasive surgery, US imaging has received significant attention in recent years because of its advantages of being widely available, non-ionizing and real-time performance \cite{douglas2001ultrasound}. Furthermore, the US offers the unique benefits of a wide range of transducers that can be used in different application scenarios from the operation room to emergency medical units \cite{scanlan2001invasive}. As a consequence, US-guided interventional procedures have been investigated and applied in different clinical applications, such as biopsy \cite{hatada2000diagnostic,coplen1991ability}, regional anesthesia \cite{barrington2013ultrasound}, ablation therapy \cite{sheafor1998abdominal,machi2001ultrasound}, prenatal diagnosis and therapy \cite{oepkes2007successful} and cardiac interventions (structural and congenital heart disease). 

\begin{figure}[tb]
\centering
\includegraphics[width=9cm]{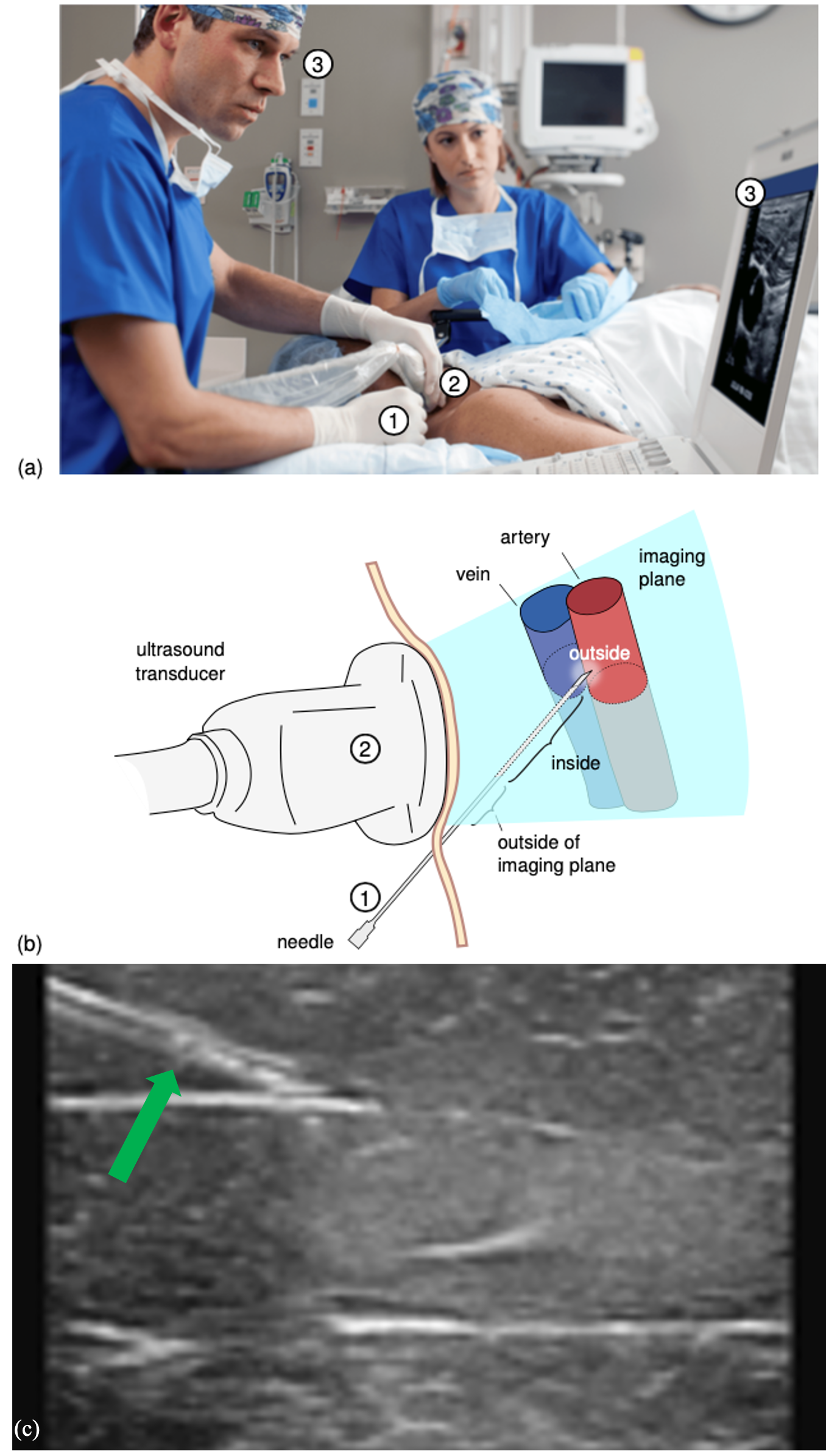}
\caption{US-guided needle therapy \cite{arashthesis}: (a) the clinical staff has to manage the multi-fold coordination of $\textcircled{1}$ the needle, $\textcircled{2}$ ultrasound transducer, while $\textcircled{3}$ looking at the US screen (Courtesy of Philips Ultrasound). (b) Schematic representation of guiding a needle using US imaging, depicting an example situation for regional anesthesia, where the needle tip is outside the imaging plane and is approaching an erroneous target area. (c) B-mode US slice contains the needle, pointed by green arrow.}
\label{fig1}
\end{figure}

\begin{figure}[tb]
\centering
\includegraphics[width=9cm]{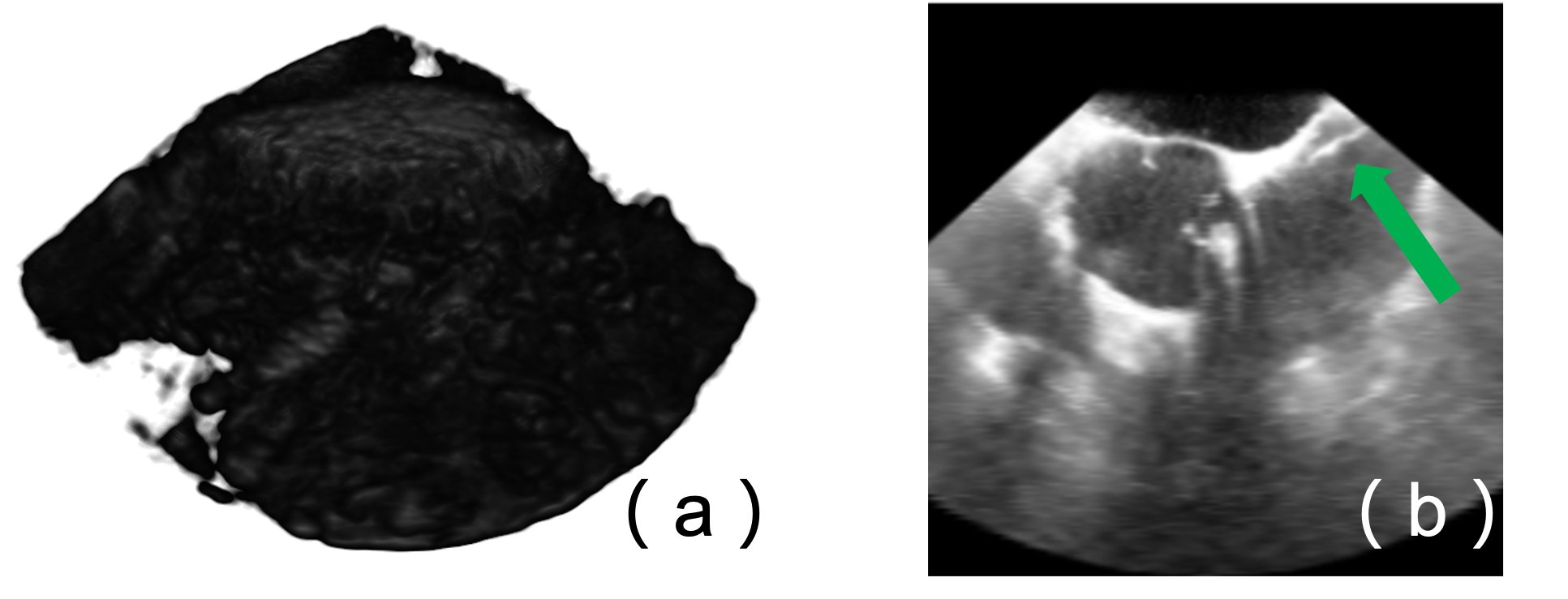}
\caption{Example of 3D US imaging in cardiac operations. (a) A rendered 3D volumetric data, where it is hard to interpret the image and localize the instrument in the volume. (b) The manually sliced b-mode image from 3D volume, which contains the instrument for cardiac operation (pointed by green arrow).}
\label{fig2}
\end{figure}

In terms of US imaging, there are commonly two formats being used, i.e., 2D images or 3D volumes, which are facing different challenges regarding instrument detection in clinical practice. As for 2D images, the multiple coordinates alignment is the most challenging part for sonographer. For example, in the needle-based anesthesia, as shown in Fig.~\ref{fig1}, to visualize the instrument in US imaging, the US plane needs to be placed perfectly parallelized to the needle. Thus the clinical staff has to carefully align the coordinates between the needle and US probe while looking at the US screen for the operation. Besides the above alignment difficulties, it is still challenging for sonographers to distinguish the instrument from the background tissue in the B-mode images by human eyes, which requires extra training to achieve the instrument interpretation, such as Fig.~\ref{fig1} (c). The 3D US imaging format is demonstrated by a rendered 3D volumetric data on the console screen, which is hard to be interpreted directly. As a consequence, a common practice is to manually extract the slice containing the instrument from the volumetric data, which is time-consuming and unfavorable for clinical practice. An example is shown in Fig.~\ref{fig2}, which demonstrate a rendered 3D volume and a sliced plane contains the instrument in the cardiac operation. Even the slice is obtained, it is still difficult to localize the tiny instrument for the operation by human eyes. Therefore, for the US-guided minimally invasive interventions, automatic instrument detection and tracking could reduce procedure time, in both 2D and 3D US imaging for image interpretation and instrument localization. It would simplify the manipulations of the ultrasound transducer, which reduces the time for the sonographer to identify the instrument inside the body and perform the intervention. Therefore, the total complexity is reduced with a simplified procedure, which is beneficial to both patient and interventionalist, especially for the therapies with exposure to radiation, like cardiac or vascular catheterization.

\begin{figure}[tb]
\centering
\includegraphics[width=9cm]{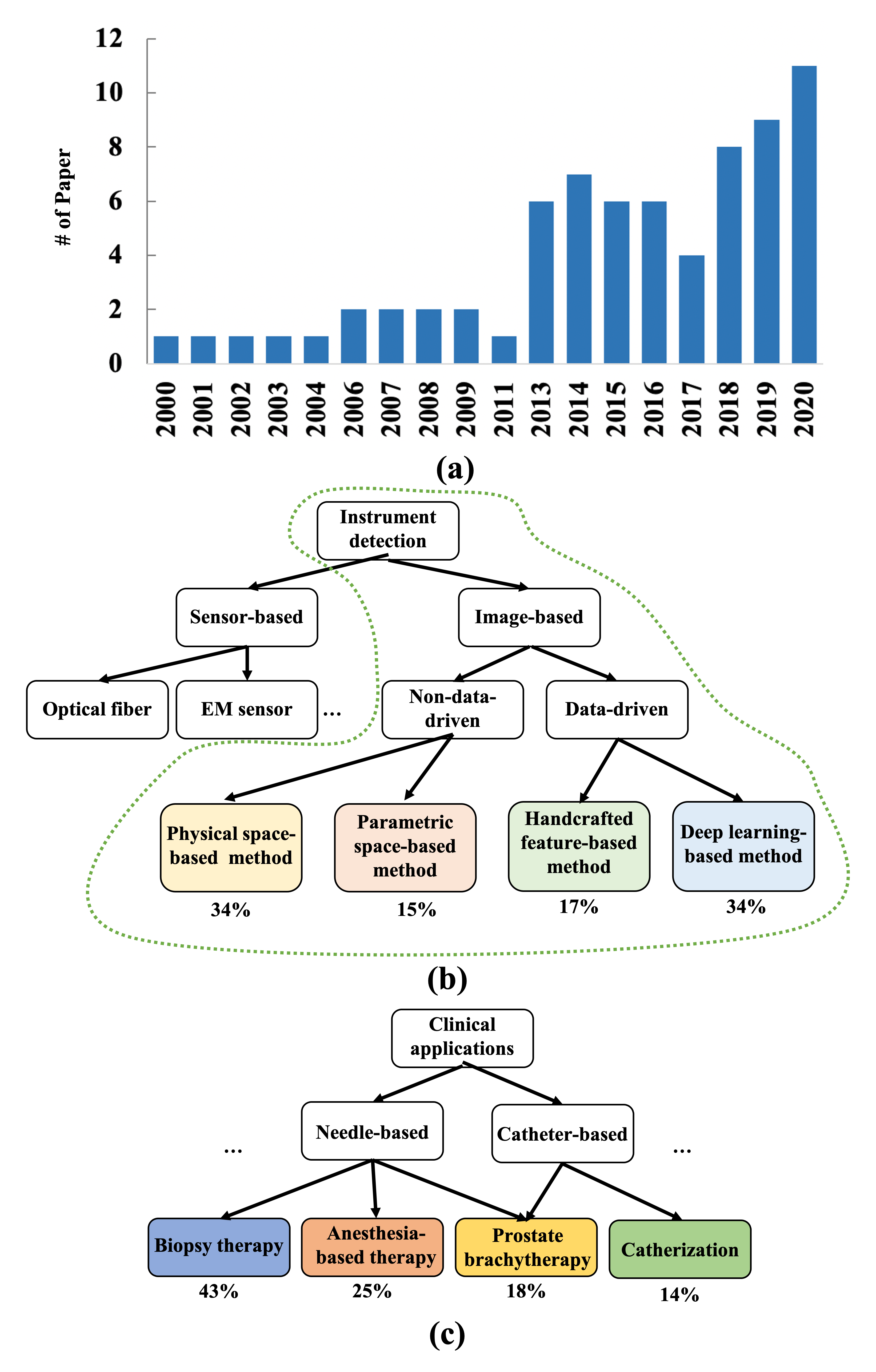}
\caption{Breakdown of the papers included in this review according to (a) the year of publications until the beginning of 2021, (b) instrument detection methods (Section~\ref{Appraoches}, this article focuses on the region in green dash line), and (c) major clinical applications (Section~\ref{Clinical}). }
\label{fig3}
\end{figure}

Existing approaches for instrument detection can be classified into two types: 1) instrument detection based on external or internal sensing devices (hardware-based), such as optical fiber sensing \cite{xia2015plane}, electromagnetic tracking \cite{krucker2007electromagnetic}, and robotic-guided detection \cite{nadeau2014intensity}; 2) image-based approaches, without using any additional sensors or devices. Although sensing-based methods have achieved promising results, their relatively high cost of equipment and the involved sensors complicated the system set up in the operation room. Therefore, a broad acceptance of the sensing-based approaches is hampered in clinical practice. In contrast, image-based approaches have been proposed to detect the medical instrument in US images. Starting with the instrument modeling by simple ultrasound image intensity analysis in work published by \cite{draper2000algorithm} in 2000, and up to the latest developed deep learning-based segmentation, various approaches have been introduced. As illustrated in Fig.~\ref{fig3} (a), the first work on medical instrument detection appeared in 2000, though the number of papers grew rapidly after 2012. This paper presents a comprehensive review on image-based instrument detection in ultrasound for minimally invasive interventions. To the best of our knowledge, this is the first review paper on medical instrument detection in US imaging. Recently Beigi \emph{et al.} \cite{beigi2020enhancement} published a review on needle localization and visualization in US imaging; in contrast, our paper has a more broad scope, covering the medical instrument detection in general. 

The review article is organized as follows. Section~\ref{Appraoches} provides the review of the instrument detection methodologies, which clusters the related literature into non-data-driven and data-driven approaches. More specifically, non-data-driven methods are grouped as they are based on a-priori assumptions on the medical instrument shape via the local image intensity distribution of voxels or pixels. They includes physical space-based methods and parametrical space-based methods, where they detect the instrument in normal physical space and parametric transformation, respectively. Data-driven methods are based on vastly studied machine learning approaches, where the instrument-related information can be modeled and learned from the data. The literature can be categorized into handcrafted feature-based methods and recent deep learning-based methods. The above classification is based on the main method of the literature, which can also include pre-/post-processing in the considered framework. The published papers, categorized based on the adopted methods, is summarized in Fig.~\ref{fig3} (b). Section~\ref{Clinical} discusses the main clinical applications, including anesthesia therapy, biopsy, prostate brachytherapy and cardiac interventions with catheters, of which a literature overview is summarized in Fig.~\ref{fig3} (c). Section~\ref{CommonData} systematically introduces the evaluation metrics, datasets, and experimental results of these papers. Section~\ref{Discussions} discusses the remaining key issues and potential research directions. Conclusions are given in Section~\ref{Conclusions}.
\begin{figure*}[htb]
\centering
\includegraphics[width=16cm]{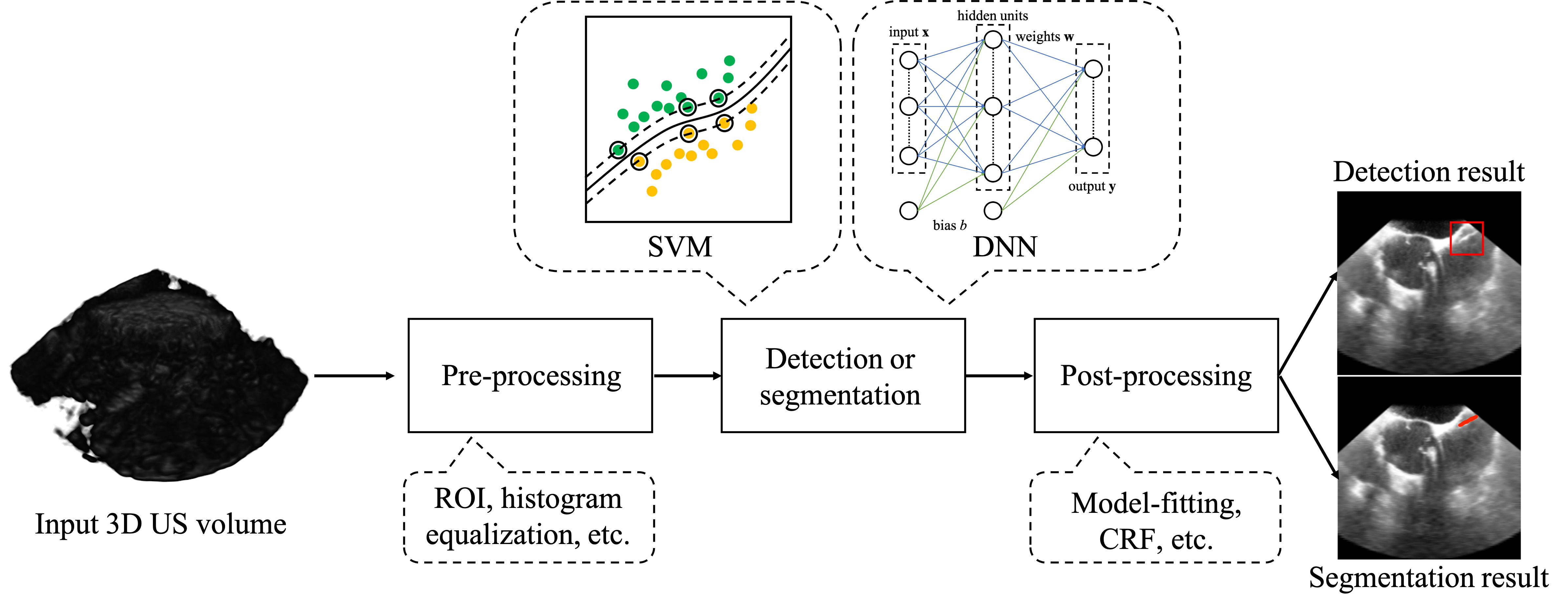}
\caption{Systematic pipeline of the instrument detection in US image. First, the input US image is pre-processed, such as histogram equalization or cropping. Then, as the key step of the pipeline, detection or segmentation algorithm is applied to localize the instrument in the image. Finally, based on the result of the detection or segmentation, the instrument is extracted from the image and is refined by the post-processing.}
\label{fig4}
\end{figure*}

\section{Instrument Detection Methods}\label{Appraoches}
The systematic pipeline of the image-based instrument detection methods is shown in Fig.~\ref{fig4}. As can be seen, the detection or segmentation algorithm is applied on the pre-processed image, such as region-of-interest cropping, image normalization or other transformations to make the images fit the next steps. Specifically, the detection is defined as an algorithm to localize the instrument in the US image with, e.g., a bounding box, axis or skeleton of the instrument. Alternatively, the segmentation algorithm is to derive a more precise result that it assigns each pixel (voxel) in the image a category, i.e., the instrument or not. Therefore, segmentation is more difficult than the detection algorithm in most cases. In this article, the detection is a more generalized term than the segmentation, since it is easy to obtain the detection results based on the segmented images. Finally, the post-processing is applied on the detection or segmentation results to refined the output of the algorithm, such as model-fitting, which improves the detection accuracy for a better instrument representation.

In this section, the detection step is considered as the key component for the instrument detection pipeline, and the existing detection methods are divided into two classes: non-data-driven methods and data-driven methods. Many of the non-data-driven publications originate from the period prior to the era of machine learning. Later on, driven by the rapid development of computing hardware and software, such as faster GPUs with larger memories, advanced US imaging systems, and open-source machine learning toolboxes, instrument detection solutions have shifted to data-driven methods, which are commonly denoted as machine learning. 
\subsection{Non-data-driven Methods}\label{NDM}
As for non-data-driven studies in instrument detection, there are two major classes based on how a-priori knowledge of the instrument is used: physical space methods and parametric space methods. The physical space method performs mathematical or geometrical modeling in the physical space with the standard spatial coordinate system in a straightforward approach. In contrast, the parametric space method detects the instrument by applying the specific spatial transformation, i.e., spatial transformation from physical space to parameter space, on an intensity image after thresholding with prior knowledge of the tool. 

\subsubsection{Physical space methods}
The preliminary study for instrument modeling can be traced back to Draper \emph{et al.} \cite{draper2000algorithm}, Smith \emph{et al.} \cite{smith2001three} and Novotny \emph{et al.} \cite{novotny2003tool} with coarse segmentation. They modeled the instrument as a regional voxel cluster with the longest and straightest connectivity groups, which was implemented based on group connectivity analysis or Principal Component Analysis (PCA). After these initial studies, a texture-based instrument segmentation method \cite{linguraru2006texture} was proposed for 3D US images by applying Expectation-maximization, local texture analysis, and PCA, which iteratively segments the instrument from an \emph{in-vitro} dataset. Zhao \emph{et al.} \cite{zhao2009needle} and Qiu \emph{et al.} \cite{qiu2014phase} introduced 3D gradient orientation to calculate the instrument phase information, which segments the needle by applying Line-Support-Region analysis for grouped regions. Similar to this local gradient analysis, a histogram analysis method was designed by McSweeney \emph{et al.} \cite{mcsweeney2014estimation} to threshold the US image, which localizes the needle by morphological operation and line fitting. The Frangi vesselness filter \cite{frangi1998multiscale} was considered to better describe the instrument and filter out instrument-related points in US images, based on assumptions on the high contrast of instrument edges compared to the background and a tubular structure \cite{ren2011tubular,mohareri2013automatic,zhao2013automatic,malekian2014noise}. Although these methods may include different pre-/post-processing steps, the core idea is to extract tubular-like structures by Hessian matrix analysis for local intensity distributions. Further processing steps range from a simple thresholding \cite{ren2011tubular,mohareri2013automatic,agarwal2019real} to Random sample consensus (RANSAC) model-fitting \cite{malekian2014noise} with Kalman filtering in time sequence-based US datasets \cite{zhao2013new,zhao2013automatic}. Moreover, the local Hessian matrix is also applied to detect the shadow of steep needles in 3D US \cite{pourtaherian2016automated}, which automatically extracts the 2D slice containing the needle for in-plane visualization. 

Besides the above vesselness filter-based methods, template matching with a pre-defined catheter filter was proposed by Cao \emph{et al.} \cite{cao2013automated} for coarse segmentation, which considered a 3D catheter template for candidate voxel selection. The resulting images were optimized by a likelihood map with shape measurement. Similarly, automatically optimized Gabor filter methods \cite{kaya2014gabor,kaya2014needle,kaya2015real,hacihaliloglu2015projection,mwikirize2016enhancement} were used with different image processing steps for needle segmentation in 2D US. Specifically, Kaya \emph{et al.} \cite{kaya2014gabor,kaya2014needle,kaya2015real} proposed to employ a two-stage method for needle localization based on Gabor filtering with an optimized insertion angle estimation. First, the Otsu's method is applied to obtain the binary image. Then, the needle in the binarized image is localized by RANSAC model-fitting, which generates the region-of-interest (ROI) for needle-tip probability mapping and localizes the tip. Their methods were validated with static images \cite{kaya2014gabor} and real-time video \cite{kaya2014needle}. They further implemented a simulation platform for needle tracking \cite{kaya2015real} for real-time localization. In contrast to Kaya \emph{et al.} methods with complex post-processing, Hacihaliloglu \emph{et al.} \cite{hacihaliloglu2015projection} employed log-Gabor filters to extract phase-symmetry information, which automatically selects the scale, bandwidth, and orientation parameters to enhance the contrast of the needle. The needle is then detected by a modified Maximum Likelihood Estimation SAmple Consensus (MLESAC) method \cite{torr2000mlesac}. Furthermore, Mwikirize \emph{et al.} \cite{mwikirize2016enhancement} proposed to localize the needle by introducing signal transmission maps for the 2D US, which firstly enhances the visibility of the needle in noisy US images. The needle is then localized by applying the algorithm from Hacihaliloglu \emph{et al.} \cite{hacihaliloglu2015projection}. 

In contrast to the above methods for static US images, some papers have been focusing on exploiting temporal information. Kaya \emph{et al.} \cite{kaya2016visual} proposed to track the needle tip by applying a dynamically updated template to 2D US video, which measures the similarity between the template and US images to identify the target. This method avoids needle localization in each video frame but requires a defined template. Beigi \emph{et al.} \cite{beigi2015needle,beigi2016automatic,beigi2016spectral} intensively studied needle detection by applying spectral analysis, which makes use of spatiotemporal information from natural hand tremor. This periodic pattern is hardly observed by human eyes but can be captured by spectral analysis of 2D B-mode images, which leads to a better result than static images. However, these image modalities were only applied to 2D+T format due to hardware constraints and complex filtering steps or real-time requirements for 3D imaging. Although there is a recent study focusing on 3D volumetric data with temporal information \cite{daoud2018hybrid}, they considered an extra camera for giving support information.

The above methods mainly follow an algorithm pipeline, which we denote as \emph{segmentation-modeling} pipeline. First, carefully designed filters or instrument templates are applied to extract or enhance the instrument-related information in the image. The optimized thresholding is then applied to binarize the images to coarse segment the instrument from the data. Second, post-processing method, such as model-fitting algorithms in 2D/3D images are applied to localize the target. Although the processing steps can be different, most of above methods indicated that a successful segmentation method is the key step in detecting the instrument in challenging US images, which heavily depends on the first steps. However, these segmentation methods are limited by prior knowledge of the instrument and are sensitive to image modality or appearance. Moreover, simple thresholding with prior or empirical knowledge also limits the segmentation performances in different application cases. To better describe the instrument-related information and obtain more accurate segmentation results, data-driven methods have been exploited to better describe the instrument with knowledge learned from the data.

\subsubsection{Parametric space methods}
Besides the straightforwards studies in a common physical space, other explorations in space transformations were studied, which is denoted as parametric space transformation. The methods in this category apply a spatial parametric transformation on US images with prior shape knowledge of the instruments, assuming e.g. a straight or curved line in the 2D or 3D space. With the prior shape information, the instrument has a strong response after some projection-based spatial parametric transformation, which accumulates the pixel or voxel intensities along with the instrument propagation with respect to spatial location and direction. Here, it is assumed that the instruments yield a higher intensity value than the background in the B-mode or thresholded US images. Ding \emph{et al.} \cite{ding2002automatic,ding2004projection} proposed to segment a needle in the 3D US by applying spatial projection from 3D to 2D thresholded volumetric data, which iteratively adjusts the projection direction in 3D space to minimize the projected needle area in the 2D plane.

Okazawa \emph{et al.} \cite{okazawa2006methods} introduced the use of the 2D Hough Transformation with prior knowledge of the needle insertion angle. The pixel values along the estimated direction are accumulated to generate a histogram of the projected voxels, which produces the corrected needle direction with post-processing, including iteratively direction rotation, points rejections, etc. Zhou \emph{et al.} \cite{zhou2007automatic,zhou2008automatic} and Qiu \emph{et al.} \cite{qiu2013needle} presented the 3D Hough Transformation-based methods on a thresholded volumetric image, which select the highest accumulated values in the transformed space as the spatial parameters of the needle. Similarly, Radon Transformation-based Parallel Integral Projection \cite{barva2008parallel} on the voxel intensity was introduced by Barva \emph{et al.} to detect the straight electrode in 3D US images. The instrument is detected as the maximized response point in the Radon parametric space, which accumulates the voxel intensity values along with the propagation of the instrument. However, the essential insight of their methods is similar to the Hough Transformation-based approaches, except for the case of thresholding. Later on, the Hough Transformation method was also applied on the 2D images by projecting 3D images using a ray-casting approach \cite{aboofazeli2009new}. Recently, Beigi and Rohling have employed temporal information to enhance the ability of the Hough Transformation \cite{beigi2014needle}, which detects the needle in the 2D+T(time) US images. Daoud \emph{et al.} \cite{daoud2018accurate} also applied the 2D Hough Transformation technique to needle localization in 2D B-mode images, where they introduced Power Doppler as supporting modality to improve the performance. 

The above methods have some prior knowledge or requirement for instrument detection: (1) instruments are straight or a little bit curved in the images, which can be modeled by accumulating the values along the instrument axis through a spatial transformation; (2) the instrument has higher intensity values than the background such that a simple threshold or intensity-based transformation can be directly applied to detect the target. With the previous assumptions, most methods were validated based on computer simulations \cite{barva2008parallel} or phantom environment (\emph{in-vitro}) \cite{ding2002automatic,okazawa2006methods,zhou2007automatic,zhou2008automatic,qiu2013needle}; the challenges and difficulties were underestimated for instrument detection in noisy B-mode images. Although there are some studies \cite{ding2002automatic,zhou2007automatic,qiu2013needle,beigi2014needle} considering more challenging datasets from isolated tissue or even patient data, i.e., \emph{ex-vivo} and \emph{in-vivo} datasets, the simplified voxel thresholding without considering sufficient local or contextual information hampers the capacity of the detection algorithms, leading to many outliers or an under-segmented instrument. In contrast, instrument model-fitting methods with proper pre-processing, such as a better segmentation, could be more suitable to model the instrument in B-mode images and improve the detection performance \cite{zhao2017evaluation}.  

\subsection{Data-driven Methods}\label{DM}
\begin{figure*}[htbp]
\centering
\includegraphics[width=14cm]{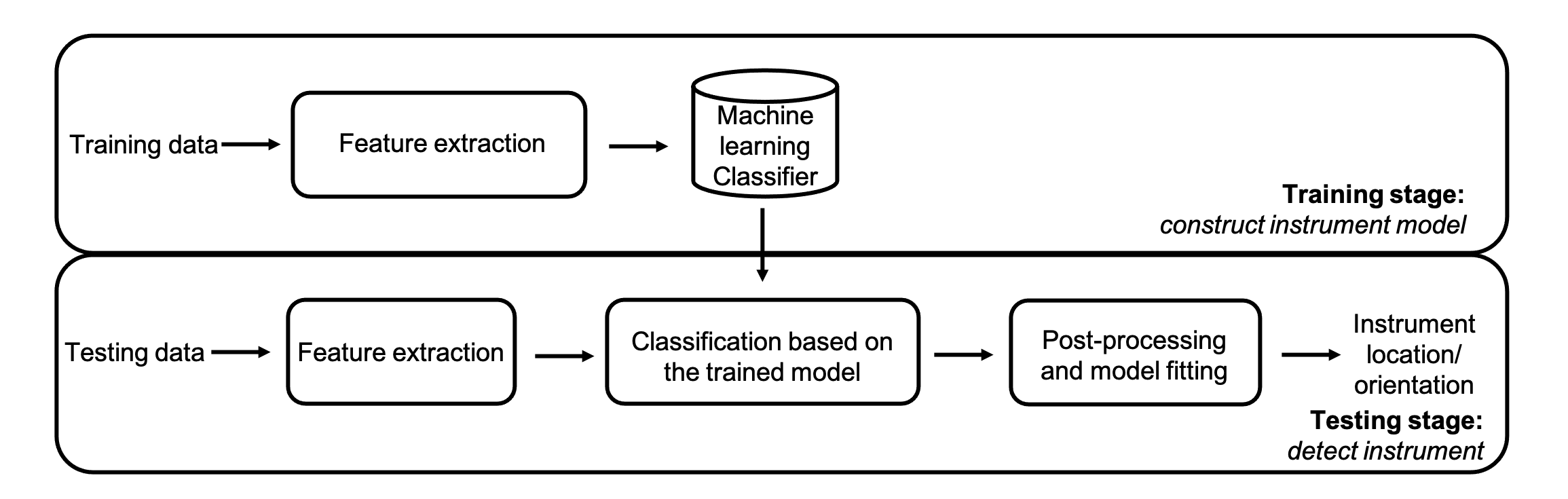}
\caption{A systematic overview of the machine learning based instrument detection pipeline. During the training stage, the machine learning model is constructed based on the training data. During the testing, the feature vector of the input data is extracted and classified by the pre-trained model.}
\label{fig5}
\end{figure*}

In recent years, with the fast development of hardware for data recording and processing, data-driven methods, also called machine learning methods, have been intensively exploited in computer vision and medical image analysis areas. The main idea of data-driven methods is to model task-related information by designing a proper mathematical representation from the training dataset, e.g., feature vector and pre-trained classifier, which is then used to make a prediction or decision. There are two popular and widely studied approaches. The first one is handcrafted feature design with a machine learning classifier, which is depicted in Fig~\ref{fig5}. This method employs feature vector extraction and task classification, which are commonly applied at voxel-level for the instrument segmentation. Specifically, the \emph{segmentation-modeling} pipeline is reformulated, which applies the machine learning-based voxel-level classification to have a better instrument segmentation result. However, the design of handcrafted features requires task-related knowledge and experience, which hampers the classification performance. Therefore the traditional machine learning approach is gradually replaced by the recently developed deep learning technology. Deep learning is a fully data-driven method, which combines feature extraction, training and classification with a fully automated information learning style. Deep learning methods can automatically learn the task-related information from provided data and, in most cases, learn more powerful representations than the handcrafted feature design methods.

\subsubsection{Handcrafted feature-based methods}
\begin{figure*}[htbp]
\centering
\includegraphics[width=15cm]{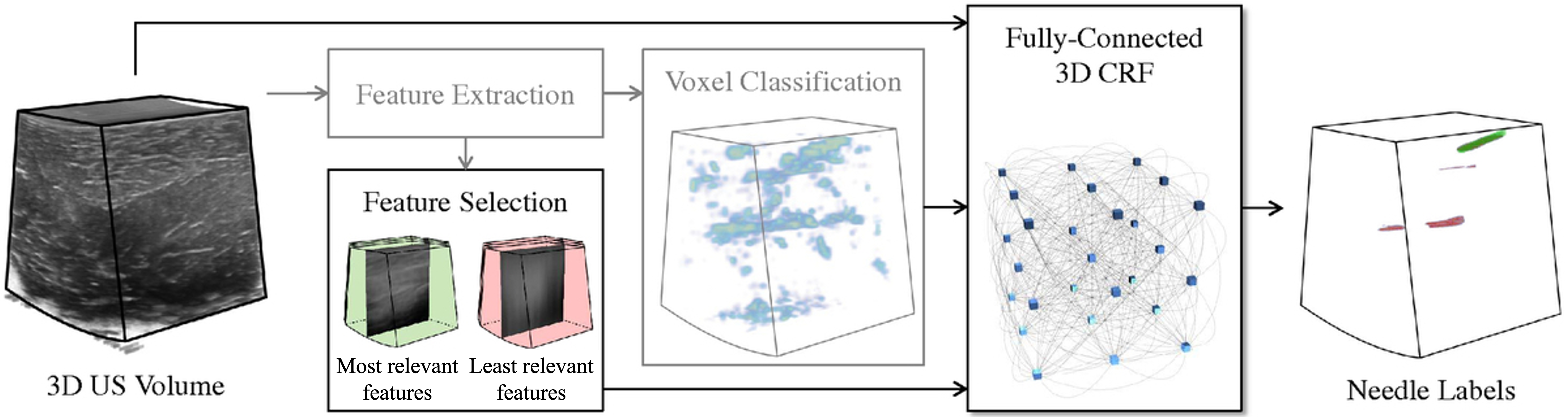}
\caption{Block diagram of CRF-based needle detection \cite{zanjani2018coherent}. The features for each voxel are extracted for classification. Besides, the feature selection is applied on the feature vectors for fully-connected 3D CRF, which is combined with original image and classified volumetric data. With the analysis of inputs and spatial correlations, the 3D CRF processes contextual relationships between selected features and the segmented volume.}
\label{fig6}
\end{figure*}

Because machine learning methods provide better instrument segmentation performance \cite{krefting2007segmentation,uhervcik2013line}, they have been intensively studied in recent years. Uher{\v{c}}{\'\i}k \emph{et al.} \cite{uhervcik2013line} proposed to use voxel intensity, Frangi vesselness response and axis descriptors as the discriminating features, which are categorized by a cascade classifier for needle detection. Specifically, cascade classifier is a type of meta-learning method to sequentially combine several different machine learning classifiers to gradually refined the classification results. Compared to the non-data-driven methods, such as Parallel Integral Projection and Random Hough Transformation, the voxel-level classification provides a higher detection accuracy because of a better segmentation result can be obtained. Similar to Uher{\v{c}}{\'\i}k, Hatt \emph{et al.} \cite{hatt2015enhanced} proposed to describe the pixels in 2D US images with the second-order Gaussian derivative filters, which are classified by AdaBoost for segmenting the needle. Their results demonstrated their method achieved better performance than the straightforward thresholding methods like intensity thresholding, filtered thresholding, or Frangi vesselness thresholding. 

Pourtaherian \emph{et al.} extensively studied needle segmentation by applying 3D orientation-invariant Gabor features with Linear Support Vector Machine (LSVM) \cite{suykens1999least}, where the segmentation results are further processed by a RANSAC algorithm to localize the instrument \cite{pourtaherian2015benchmarking,pourtaherian2015multi,pourtaherian2017medical}. Specifically, the Gabor filter bank with a pre-defined scale and multiple orientations is applied on the 3D volumes. For each voxel, the corresponding filter responses are extracted as the feature vector, which is then circular shifted for orientation-invariant. The voxel is classified by the SVM classier to determine whether it is needle or not. With classification on the all voxels of the volume, the 3D US is segmented for further post-processing, e.g., RANSAC model-fitting. Their studies show that the Gabor filter with LSVM could adequately capture the spatial information for a long but thin instrument in complex 3D US images. Later on, Zanjani \emph{et al.} \cite{zanjani2018coherent} demonstrated the Gabor features could be simplified by feature selection and the segmentation performance can be boosted by Conditional Random Field (CRF) \cite{lafferty2001conditional} compared to a simple LSVM, which segments the images based on contextual-level information correlation (as shown in Fig.~\ref{fig6}). In more details, the voxel's spatial position is jointly correlated with selected feature vector, which exploits the contextual information of the instrument based on the segmentation results from LSVM classifier. 

Mwikirize \emph{et al.} \cite{mwikirize2017local} also considered log-Gabor features as the local phase extractor, which is processed by a locally normalized histogram of orientated gradients (HOG) features, to describe the needle in 2D US slices. The constructed HOGs are then classified by LSVM to segment and enhance the needle in 2D US slices. In the meantime, Younes \emph{et al.} \cite{younes2018automatic} proposed to make use of a Gaussian mixture model-based classifier to segment the needle in 3D prostate brachytherapy US. The needle is finally localized by a standard RANSAC model-fitting. To further exploit complex anatomical information in the 3D US for catheter segmentation in catheterization, Yang \emph{et al.} proposed multi-scale and multi-definition features for supervised learning classifiers, which demonstrated a better discriminating information extraction than techniques solely based on Gabor features \cite{yang2018feature,yang2019catheterJMI}. The segmented instrument was fitted by a more complex Sparse-Plus-Dense RANSAC algorithm to fit the curvature instrument in the cardiac chambers. 

The above methods are applied on static images, because the voxel-level feature extraction and classification in 3D+T US images are computationally expensive for current hardware. Nevertheless, several papers were proposed to focus on 2D US with temporal information. Beigi \emph{et al.} \cite{beigi2017detection} proposed to detect a hardly visible needle in 2D+T US by applying local phase extraction with temporal sequence analysis. More specifically, the phase information for each frame is extracted to formulate the element from a time-sequence-based phase video, which is then processed by the Auto-Regressive Moving-Average (ARMA) model to extract the feature vector. With classification from a modified SVM, the small motion of the needle can be characterized for needle detection. Furthermore, Beigi \emph{et al.} employed spectral feature analysis on spatiotemporal features derived from optical-flow analysis \cite{beigi2017casper}, which allows to detect and track the needle in a 2D US video. In contrast to these off-line learning methods, Mathiassen \emph{et al.} \cite{mathiassen2016robust} applied on-line learning, i.e., learning and updating the needle-related information during the video progressing. They applied the statistical filtering methods, i.e., Kalman filter and particle filter, to learn the appearance of the instrument in the video with real-time performance.

Besides the above machine learning methods, Zhang \emph{et al.} \cite{zhang2020multi,zhang2020multiTMI} proposed to detect multi-needle in 3D brachytherapy by unsupervised sparse dictionary learning (SDL). Specifically, the needles and tissue information in the 3D images are encoded into latent space, which are then distinguished by the sparse dictionary model. Based on the SDL, the needle in the 3D space can be captured and reconstructed in the volume that multiple needles can be localized by region-of-interest-based RANSAC. 

Even though the above methods achieved promising detection results for the given tasks, it is challenging to design the optimal feature representation, which is one of the key factors that hampers the segmentation performance. Because of this, the complex post-processing is needed to avoid outliers or false positives. Moreover, the designed features can only focus on local information while ignoring the contextual and semantic information \cite{zanjani2018coherent}. To handle these limitations, deep learning methods have been studied recently~\cite{litjens2017survey}.

\subsubsection{Deep learning based methods}
At the beginning of the deep learning era, Geraldes and Rocha proposed to consider neural networks, i.e., Multilayer Perceptron network (MLP), to segment the needle in 2D US images \cite{geraldes2014neural,rocha2014flexible}. The segmented results guide the Kalman filter to track the needle tip in the video sequences \cite{geraldes2014neural}. These papers demonstrated the feasibility of deep learning to detect a medical instrument in challenging US images. 

\begin{figure}[htbp]
\centering
\includegraphics[width=9cm]{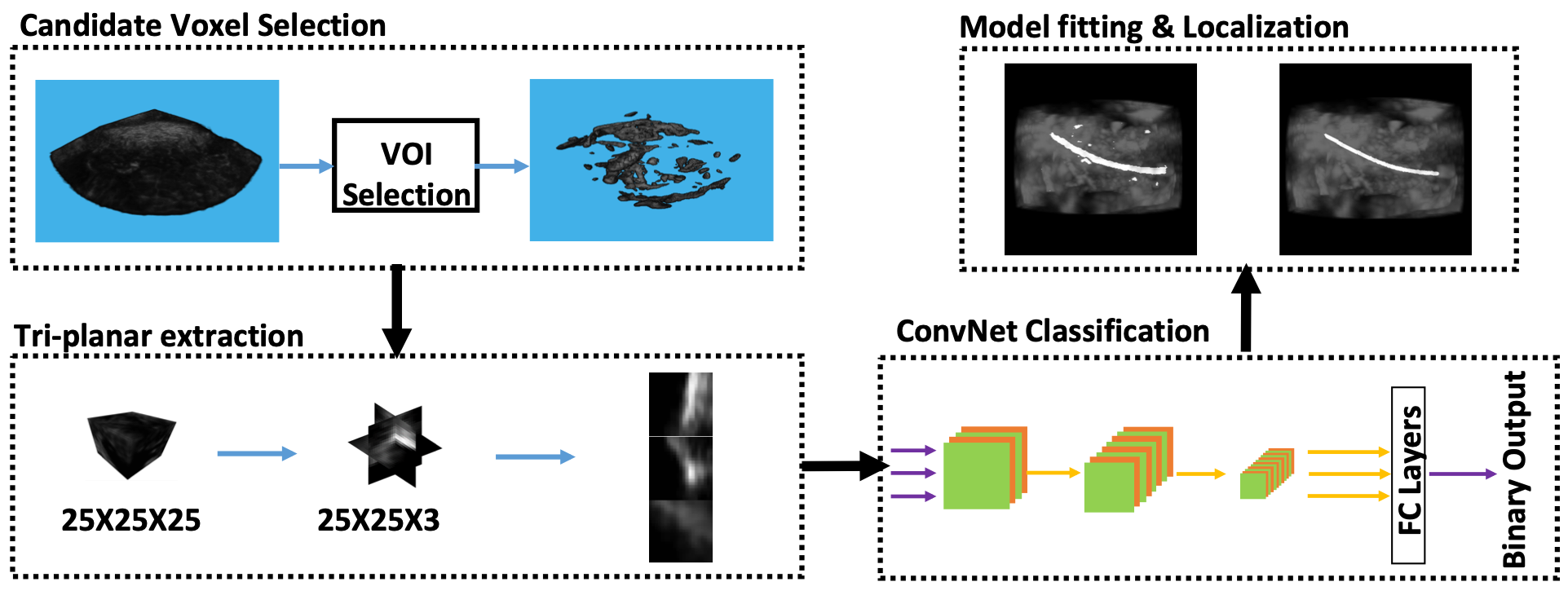}
\caption{Block diagram of VOI-based CNN for catheter detection \cite{yang2019catheter}. The input volume is first processed by a Frangi filter to select the VOI voxels, which are then classified by a tri-planar-based CNN for voxel-based classification. The RANSAC model-fitting is applied to localize the catheter in 3D B-mode images.}
\label{fig7}
\end{figure}

The conventional idea of voxel-based classification was extended into deep learning methods, which employed convolutional neural networks (CNN a famous type of deep neural networks, DNN) to replace the feature extraction and classification steps in Fig.~\ref{fig5}. The CNN model determines the voxels' category by a classification strategy on the whole 3D image. A typical approach is the tri-planar CNN method \cite{pourtaherian2017improving, pourtaherian2018robust,yang2018catheter,min2020feasibility}, which segments the instrument in 3D US volume by voxel-level classification. Specifically, for each voxel of the input image, a 3D local patch is extracted with the voxel as the center. To simplify the CNN and reduce the computational cost, three orthogonal slices passing through the voxel are extracted, which are used as the inputs for the CNN classification. The volume is segmented by classifying all the voxels. Compared to conventional handcrafted feature based methods, this deep learning method automatically learns the discriminative representation based on the training data, which therefore better exploit the information for classification. Nevertheless, the exhaustive strategy is time-consuming because it applies a CNN on every voxel of the image. Later on, this exhaustive strategy was overcome by applying the Frangi vesselness filter as a voxel-of-interest (VOI) pre-selection step \cite{yang2019catheter}, which is shown in Fig.~\ref{fig7}. A fast region-based CNN (Fast R-CNN) is combined with a regional proposal network (RPN) to efficiently detect the needle in 2D US images \cite{mwikirize2018convolution}. Specifically, the Fast R-CNN is considered to generate the shared feature maps for RPN, which classifies and regresses the location of the needle in the input image. By doing so, the location of the needle in 2D US images is annotated by a bounding box. However, this method cannot accurately segment and localize the instrument skeleton at the pixel or voxel level. To overcome this limited performance and leverage the powerful fully convolutional network (FCN), which assigns the class categories to all the points of the input image by using semantic information, semantic segmentation was introduced and studied for instrument segmentation. 

\begin{figure}[htbp]
\centering
\includegraphics[width=9cm]{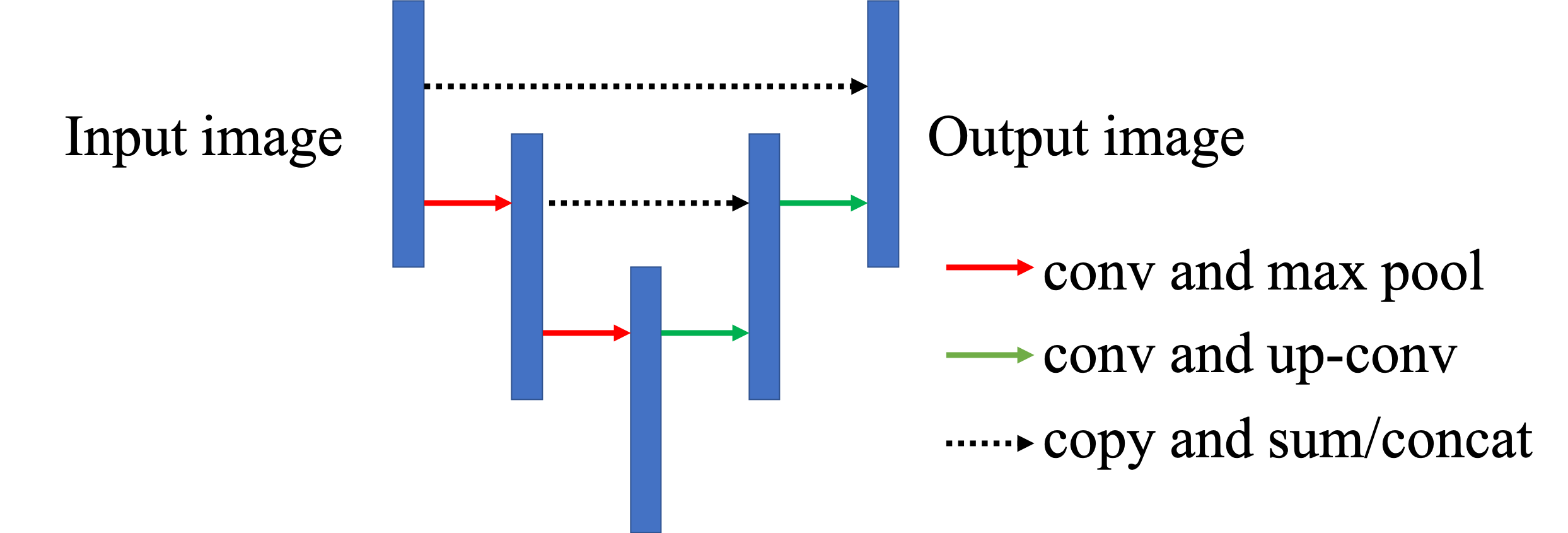}
\caption{A systematic example of U-Net structure. The U-Net include an encoder and a decoder. The input image is encoded by several convolutional and max pooling operations. The the intermediate feature maps are processed by the decoder via several convolutional and up-convolutional operations. The encoder feature maps are skipping connected with decoder to preserve the low-level information.}
\label{fig8}
\end{figure}

To semantically segment the instrument in the US, an FCN with U-Net structure \cite{ronneberger2015u} (shown in Fig.~\ref{fig8}) was considered as a solution, since it exploits the semantic information at different image scales with skipping connections for data flow. Meanwhile, the low-level information is combined with high-level semantic information, therefore, a better semantic segmentation can be achieved via end-to-end learning manner. This approach leads to state-of-the-art performance in most applications in the medical imaging area. A 2D FCN is applied to segment the needle in 2D US \cite{lee2020ultrasound,rodgers2020automatic,gillies2020deep} and in 3D US where the volume is decomposed into stacks of slices \cite{pourtaherian2018localization,yang2019efficient} for segmentation with the U-Net. However, for 3D volumetric data, the decomposition approach may limit the semantic information usage due to the compromised 3D information after slicing. To address this, patch-based 2.5D or 3D U-Net were proposed to segment the cardiac catheter \cite{yang2019ISBI,yang2019automated,yang2019MICCAI,yang2020deepQ,yang2021efficient} or prostate needles \cite{zhang2020multiMP,andersen2020deep} in 3D volumetric data by dividing the image into smaller patches or reducing the image size. In this way, the 3D contextual information is preserved and the requirement on GPU memory is reduced for 3D deep learning. Nevertheless, this patch-based strategy limits the whole image contextual information usage. To overcome this limitation, Arif \emph{et al.} applied an extremely simplified 3D U-Net on a complete 3D image to segment the needle in ultrasound imaging for liver~\cite{arif2019automatic}. Although their method showed a promising results, its generalization and segmentation abilities are constrained by the simplified network design \cite{yang2019MICCAI}. Alternatively, a multi-dimensional FCN was proposed to decompose the high-level 3D feature maps into 2D space, which simplifies the decoder and reduces the GPU usage \cite{yang2020efficient}.

Due to the requirements of a large number of training images and high GPU memory usage for deep learning, temporal information is not widely investigated in deep learning-based medical instrument detection. Mwikirize \emph{et al.} \cite{mwikirize2019single,mwikirize2019learning} proposed time-difference-based regression and classification CNNs to detect the needle in 2D US sequences. The differences between two adjacent frames are obtained by applying pixel-wise logical operation, which captures the subtle motion of the needle and feeds it into CNNs for detection. Nevertheless, these methods process the temporal information outside the CNN such that the deep learning approach may not properly handle the spatial-temporal information. 

In addition to commonly used CNNs, convolutional dictionary learning was proposed by Zhang \emph{et al.} \cite{zhang2020weakly} as an extension of their work \cite{zhang2020multi}. Instead of sparse dictionary learning (SDL), they considered a convolutional sparse coding model to replace the SDL method \cite{zhang2020weakly}, which used CT images as the supervisory signal to create the dictionary for reconstructing the detected needles in the 3D volumetric data.

Although the deep learning methods provide advantages of superior segmentation performance and better information description, this data-driven methods require a large amount of training data with annotations. To address this challenge, a semi-supervised learning method has been proposed by jointly considering uncertainty estimation and contextual constraint \cite{yang2020deep}. With the proposed method, the deep learning network requires much fewer annotations than the conventional supervised learning method while the segmentation results are comparable. Nevertheless, the study of this approach is still limited. Moreover, most state-of-the-art detection or tracking methods are far from real-time performance, so optimizations are still required. More specifically, real-time performance is more crucial for instrument tracking in the clinical practice, while the non-real-time performance is accepted for instrument detection in some operations \cite{zhang2020weakly,zhang2020multi}. These limitations also form key issues for employing deep learning when considering clinical applications.

\section{Clinical Applications}\label{Clinical}
This section discusses the main clinical applications related to medical instrument detection. More specifically, the four clinical applications receiving most attention in the literature are reviewed, as shown in Figure.~\ref{fig3}. 
\subsection{US-guided regional anesthesia}
Needle-based regional anesthesia or imposing blockade is essential in current clinical practice, which provides a safer and more accurate intervention for further procedures. Conventional regional anesthesia requires experienced radiologists to deliver the medicine to the correct region, which is commonly guided by ultrasound imaging since it provides a fast and convenient visualization solution for clinical experts. However, as shown in Fig.~\ref{fig1}, multi-fold coordination of the US screen, needle, and ultrasound probe complicates the procedure and hampers the operation outcomes with higher risks. As a result, extensive training for a surgeon is required to achieve a successful therapy under the guidance of the US. 

To visualize the needle during the US-guided regional anesthesia or blockade, an essential condition should be satisfied in conventional 2D US: the needle should be positioned in-plane in 2D images, where the needle is visualized as a bright line, requiring a perfect alignment between instrument and ultrasound plane \cite{mwikirize2016enhancement}. However, this 2D US-guided therapy faces the challenges of the instrument being invisible \cite{beigi2017detection}, or the instrument being out of the plane \cite{pourtaherian2017medical}. Consequently, 3D volumetric ultrasound is gradually adopted into clinical usage because it can provide richer spatial information of the needle. However, complex 3D information and complicated 3D image visualization hamper the efficiency of the radiologists when they are looking for the needle and guiding it to the target region. As a consequence, automatic needle detection is investigated to facilitate clinical interventions and improve operation outcomes. 

Automated needle detection for anesthesia or blockade has been studied. Several papers have extensively validated the technology on the clinical datasets. A needle tip localization method in the 2D US image is validated on bovine, porcine, kidney and liver datasets \cite{mwikirize2016enhancement}, which achieved a mean localization error of $0.3\pm0.06$ mm. Similarly, the Gabor-based needle detection method in 3D was validated on the patient dataset, which achieved the detection error of 0.68 mm. These results demonstrated good results from the clinical aspect, which is proven to be a promising solution to facilitate this type of regional intervention. Nevertheless, these limited off-line validations with small clinical datasets are not sufficient; further extensive clinical validation is needed. In particularly clinical trails for the on-line validation is necessary to directly validate the effectiveness of the algorithm for the operation guidance.
\subsection{US-guided biopsy}
\begin{figure}[htbp]
\centering
\includegraphics[width=6cm]{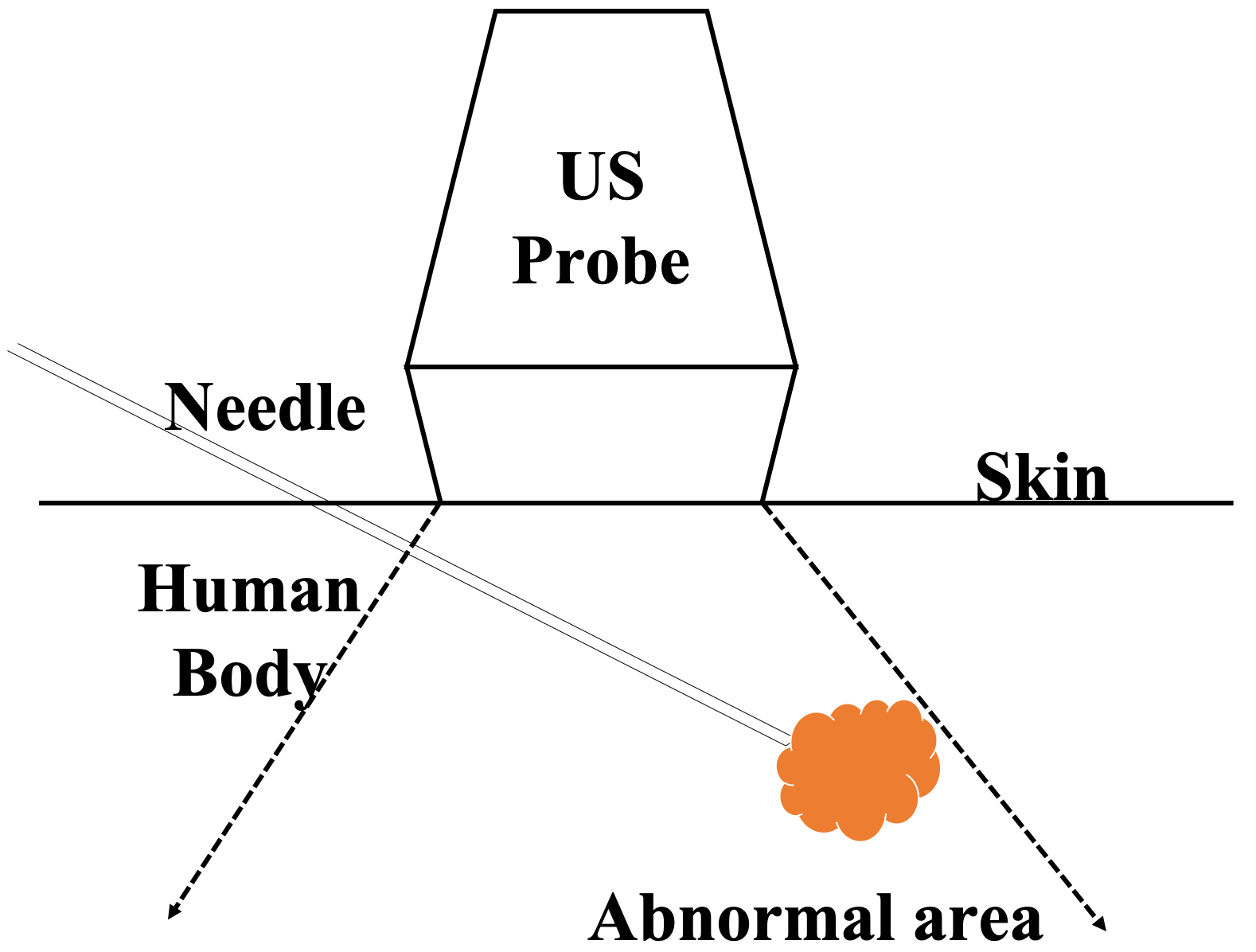}
\caption{Example of US-guided needle biopsy. With guidance of US imaging, the needle is correctly placed to abnormal regions for performing a biopsy, such as breast biopsy or thyroid nodule biopsy.}
\label{fig9}
\end{figure}

Biopsy of the nodule or lymph node is essential for diagnosis, such as liver, breast or prostate biopsy, particularly for finding malignant tissue, as shown in Figure.~\ref{fig9}. To obtain the tissue samples by biopsy taking, needle biopsy or open surgical biopsy is commonly considered, based on suspected pathology, patient health condition, and procedure complexity. Although conventional open surgery biopsy provides better diagnostic results, the less invasive needle-based biopsy becomes attractive, as it offers a reasonable result. Historically, a needle-based biopsy was performed by radiologists in special procedure rooms with interventional radiology suites, which is however gradually replaced by a US imaging system because of its lower cost, higher healthcare efficiency, and better tissue characterization. However, the drawbacks of US imaging need to be addressed, like difficult interpretation, lower image contrast than traditional X-ray imaging, and extra training required for radiologists to satisfy tissue sample \cite{patel2019ultrasound}.

Similar to the above regional anesthesia, multi-dimensional coordination has complicated the procedure of needle guidance. As a result, automatic instrument detection is necessary to help radiologists in performing their tasks. Moreover, US imaging is also increasingly used for tissue characterization or abnormality detection, facilitating the biopsy procedures and instrument detection. With a richer 3D spatial information, 3D US can efficiently and accurately facilitate the radiologists to perform the operations and reduce the risk for patients.

Arif \emph{et al.} \cite{arif2019automatic} have validated the automated needle detection on the liver biopsy dataset, which detects needle position and orientation with the mean error of 1 mm and 2$^\circ$, respectively. This experimental result seems to indicate that a robust needle detection for the 3D US-guided liver biopsy taking is possible. Nevertheless, their datasets are relatively too small for a deep learning based approach, which only includes 8-9 patient images for training and testing. Therefore, further extensive validation on large patient dataset is required for the clinical practice. In addition, biopsy taking is a common procedure for the tumor diagnostic, which can be applied on different organ or region of the patient body. Therefore, algorithm should be carefully designed to be a generalized or specified solution for different types of biopsy taking. Since the deep learning based methods can easily overfit the background anatomical tissues.

\subsection{US-guided prostate brachytherapy}
\begin{figure}[htbp]
\centering
\includegraphics[width=7cm]{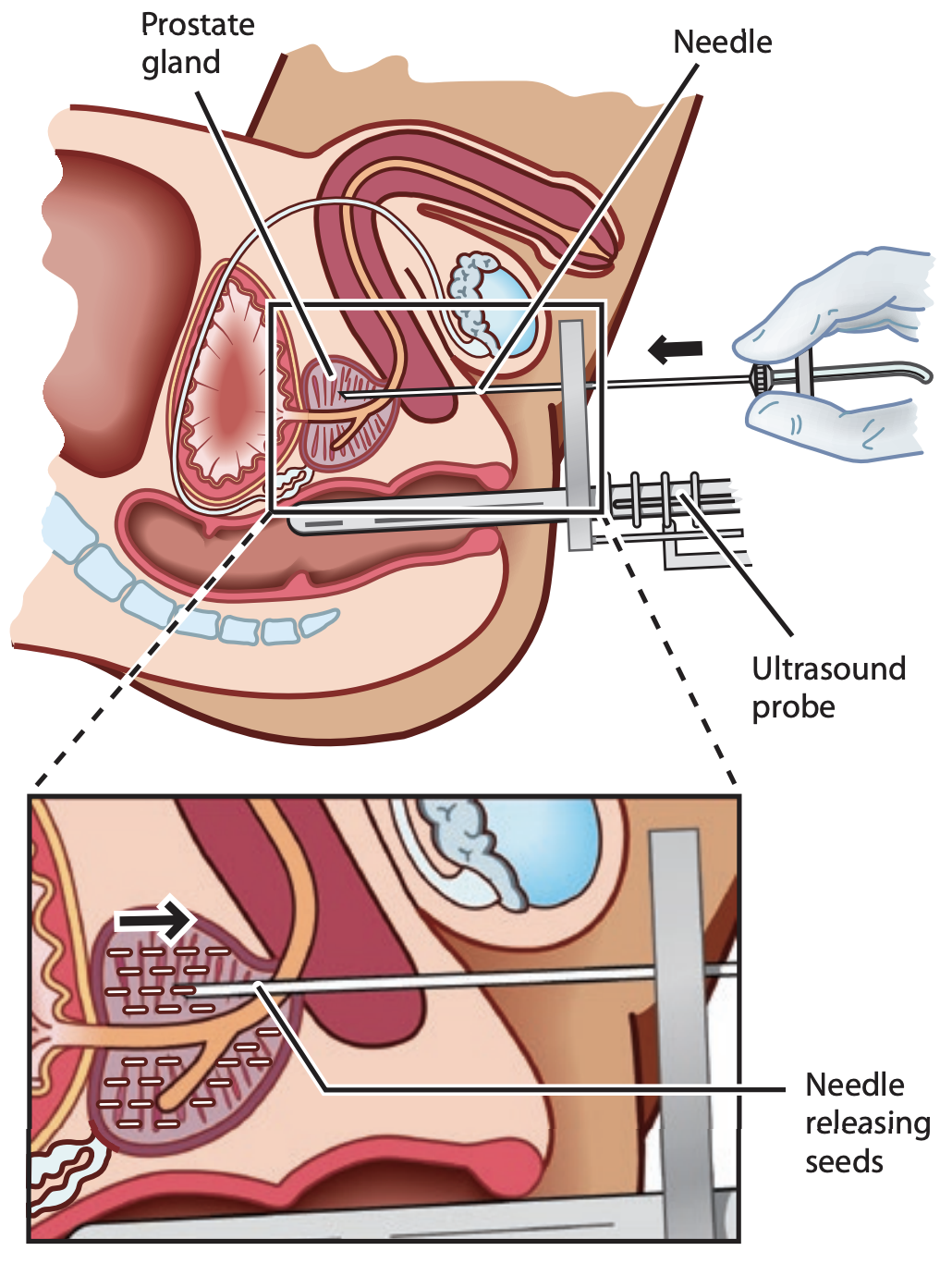}
\caption{Example of US-guided prostate brachytherapy. Needles are guided by US images to place radiation seeds for prostate cancer (source: Understanding Brachytherapy for Prostate Cancer from https://www.prostate.org.au/).}
\label{fig10}
\end{figure}

Prostate cancer is the development cancer in the prostate, which is the second-most common cancer in male patients. Prostate therapies are an important treatment worldwide for prostate cancer. Specifically, prostate brachytherapy is one of the popular treatments for patients at the early stage of cancer development. The needles or catheters are used to place radioactive particles (so-called seeds), which have the size of a grain of rice, to the tumor regions, as shown in Figure.~\ref{fig10}. It delivers high-dose radiation (HDR) to the tumor without affecting the normal tissue around abnormal areas. Transrectal Ultrasound (TRUS) has been considered since the 1980s \cite{banerjee2017use} to perform the operation, which is the key modality for the radiologist to visualize the interested region for the operation. 

In contrast to other applications, prostate brachytherapy requests to insert multiple instruments into the prostate so that multi-instrument detection in TRUS is essential for successful operation planning. However, because of this requirement and instrument placement condition, i.e., needles are close to each other, the detection algorithms need to be stable and accurate enough to detect multiple objects, which is rarely studied in state-of-the-art solutions. Moreover, besides this multi-detection challenge, an efficient detection algorithm is required because the typical operation time is around 90 minutes for prostate brachytherapy \cite{banerjee2017use}. Zhang \emph{et al.} \cite{zhang2020multiTMI} have validated the needle detection on patient dataset, which demonstrates the the instrument detection procedures can be accelerated to a half minute with tip location error around 1 mm. In contrast, conventional needle digitization takes around 15-20 minutes by an experienced physician. Therefore, automated instrument detection can facilitate the radiologists to find the instrument during the procedures.
\subsection{US-guided catheterization}
\begin{figure}[htbp]
\centering
\includegraphics[width=8cm]{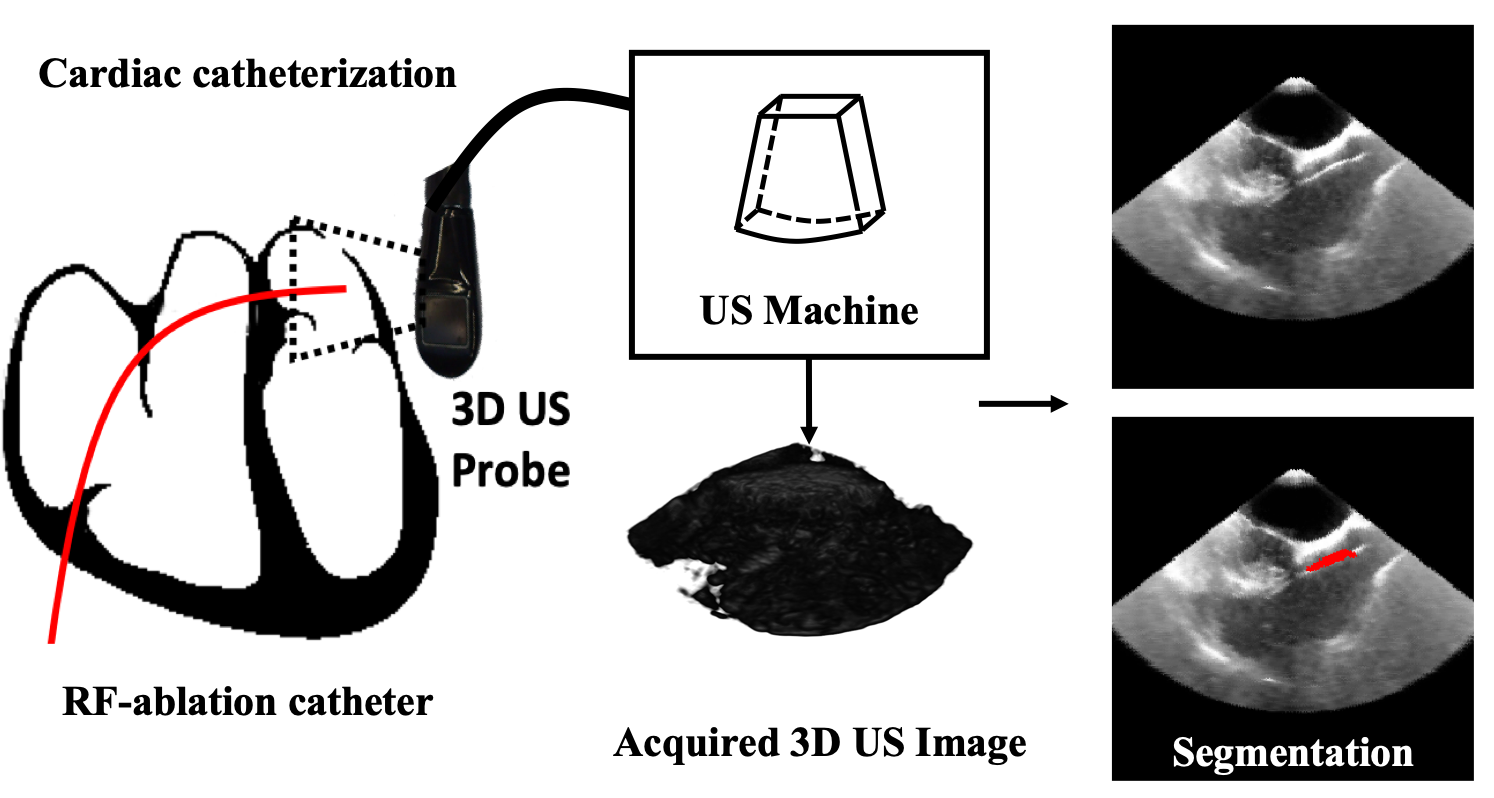}
\caption{Example of cardiac catheterization under US guidance. The ultrasound probe captures the 3D volumetric data of the heart chamber, which contains an RF-ablation catheter. The algorithm outputs the segmentation results and automatically demonstrates the results in 2D slices for a better view.}
\label{fig11}
\end{figure}
Like US-guided needle therapy, US-guided catheterizations, especially for cardiac applications like RF-ablation or TAVI procedures, face the challenge of multi-dimensional coordination for radiologists. Moreover, due to complex anatomical structures in the heart chambers, sonographers need more time to localize the less obvious catheter (compared to the metal needle) in 3D volumetric data using slice-by-slice tuning procedure, which is time-consuming and complicates the operation. Nevertheless, 3D ultrasound is attractive for catheterizations because of its radiation-free nature and easy-to-use properties, and its richer offering of spatial information for tissues \cite{yang2019catheterJMI}. As a result, it is a promising choice to support or replace current X-ray imaging for cardiac interventions. An example of cardiac catheter detection in 3D volumetric data is shown in Fig.~\ref{fig11}.

To detect the catheter in 3D US images, several solutions were proposed \cite{cao2013automated,yang2019catheterJMI,yang2021efficient}, which employed different methods as reviewed in Section~\ref{Appraoches}. However, there are still several challenges that need to be addressed in future work: (1) US-guided catheterization is not widely accepted for clinical practice, so that expert knowledge is limited by experience. Furthermore, a clear definition is lacking how accurate and how fast instrument detection should be. (2) Due to hardware limitations of the US probe, the 3D volumetric data only focuses on small field-of-view of the heart chamber, which then lacks of the guidance capability when the instrument is inserted into the target region, so that a cardiac catheter still requires X-ray to perform guidance in the vein. This hampers the flexibility of US-guided interventions. (3) Only sets of limited studies has been performed on US-guided catheterizations when compared to needle-based interventions, which makes US-guided catheterizations a minority area in the field of computer-assisted interventions.
\section{Evaluation Datasets, Metrics and Results}\label{CommonData}
\subsection{Evaluation datasets}
In this section, the papers are summarized based on the above detection methods. To complete the discussions on these papers, we summarize the information w.r.t. datasets, like data type, data format, and target clinical application, which is covered in the overview by Table~\ref{TB1} and Table~\ref{TB2}. Specifically, there are four different categories: simulation dataset, \emph{in-vitro} dataset, \emph{ex-vivo} dataset and \emph{in-vivo} dataset.

\noindent {{\textbf {{Simulation:}}}} Computer software-based simulation was used to generate ultrasound images with well-defined instrument information with image content. Field II \cite{jensen1996field} and kWave \cite{treeby2010k} are the commonly used simulation platform for US image generation. Because they can provide satisfied image results without extra cost for US equipment and experiment labs, such as a biological lab for tissue experiments and target instrument operations, which require extra clinical doctors to guide the experiments. As a result, a simulation dataset can be used to validate the scientific idea and perform a feasibility study, e.g., mimicking the needle or catheter insertion in 3D US images \cite{barva2008parallel,zhao2013automatic}. Although kWave has ability to model nonlinearities in tissue, which achieves better simulation results than Field II, this simulation approach lacks complex anatomical structures of the tissue and far from the real images from the operation, which limit the clinical value and stability of the methods. 

\noindent {{\textbf {\emph{in-vitro:}}}} In contrast to simulation datasets, \emph{in-vitro} datasets provide a more realistic case, which employ polyvinyl alcohol (PVA) or agar phantoms to mimic the human tissue \cite{draper2000algorithm,okazawa2006methods,zhao2009needle}. Moreover, these datasets were recorded from real US imaging equipment with proper post-processing, which tried to mimic real clinical scenarios. However, this approach has an essential limitation that the phantom cannot include complex and detailed anatomical structures in the clinical applications, such as vessels or muscles. Moreover, due to the different physical properties between a phantom and real tissue, the US imaging results are also different, which limits the clinical value of the algorithms. For instance, line-filtering-based needle detection algorithm was validation on a phantom dataset \cite{zhao2009needle}, which however was proven to be unstable for real tissue data \cite{pourtaherian2017medical}.

\noindent {{\textbf {\emph{ex-vivo:}}}} Similar to \emph{in-vitro} data, \emph{ex-vivo} data is another commonly used dataset type, which replaces PVA or agar phantoms to an isolated real animal tissue, such as chicken breast for needle detection \cite{pourtaherian2017medical} or porcine heart for catheter detection \cite{yang2018feature}. These tissues provide a more complex image appearance due to anatomical structures, which are much more similar to clinical applications. Nevertheless, this dataset type still has limitations in recording conditions, such as less complex muscle and vessel structure of chicken breast when considering a needle detection for anesthesia or water-filled (rather than real blood) heart chamber for cardiac catheterization. Although these limitations hamper the clinical value, they provide a more stable and promising comparison than the above non-tissue-based datasets. Because of the data recording difficulties, such as biological experiment certification and support from clinical experts, \emph{ex-vivo} datasets are still important and are considered in a majority of recent papers, which indicates their importance for algorithm validation. In addition, by comparing the performance stability, \emph{ex-vivo} dataset has more challenges than the \emph{in-vitro} images \cite{pourtaherian2017medical,yang2018feature}.

\noindent {{\textbf {\emph{in-vivo:}}}} In this article, \emph{in-vivo} data is defined as the off-line dataset coming from real clinical operation, such as prostate biopsy, or live animal dataset, e.g. live porcine for cardiac catheterization. Because the datasets were recorded from real clinical usage or mimicking it, these \emph{in-vivo} datasets demonstrated important clinical value for the algorithm validation. However, recording of this type of dataset is challenging when compared to the above three dataset types. First, it is difficult to ask radiologists to record the data when the therapy is not widely accepted in the hospital, or data recording can hamper the operation procedures. Second, due to privacy policy, it is difficult to obtain the patient dataset without complex anonymization steps. Third, the cooperation with clinical experts can be difficult, because various radiologists have different preferences for image appearance and configurations, which makes it complicated to agree on a unified image quality and appearance for algorithm.

As summarized in Table~\ref{TB1} and Table~\ref{TB2}, only the most valuable clinical datasets are reported, i.e. \emph{in-vivo} or \emph{ex-vivo} datasets, and are indicated for multi-dataset validation papers. As can be observed, the non-learning-based methods are mostly validated on less complex datasets, e.g., simulation and \emph{in-vitro} datasets. The key reason is these methods are proposed before the era of machine learning, which mostly performed feasibility studies to validate the idea of automated instrument detection. With fast development of machine learning in healthcare area, clinical datasets, such as animal tissue or patient dataset, are more easier to access for algorithm validation. Therefore, learning-based methods are mostly validated on \emph{ex-vivo} and \emph{in-vivo} datasets, which obtained promising results for clinical practice. The key reason is that the learning-based method tends to learn the anatomical structures with less generalization for different clinical applications. Therefore, a carefully prepared dataset is more meaningful for the algorithm validation, since the domain gap between technical experiments and clinical practice should be as small as possible.

\begin{table*}[tb]
\centering
\caption{Overview of papers using non-learning-based techniques for instrument detection, ordered by year of publication. Clinical applications are mentioned in the literature.}
\label{TB1}
\begin{tabular}{llllll}
\hline
Reference& Year&Format&Application&Dataset\\\hline
Draper \emph{et al}. \cite{draper2000algorithm}& 2000& 2D & needle biopsy/brachytherapy&\emph{in-vitro}\\
Smith \emph{et al}. \cite{smith2001three}& 2001& 3D & needle biopsy&\emph{in-vitro}\\
Ding \emph{et al}. \cite{ding2002automatic}& 2002& 3D & needle biopsy/brachytherapy&\emph{in-vitro/ex-vivo}\\
Novotny \emph{et al}. \cite{novotny2003tool} & 2003 & 3D &graspers for cardiac/fetal surgery &\emph{ex-vivo} \\
Ding \emph{et al}. \cite{ding2004projection}& 2004& 3D & needle biopsy/brachytherapy&\emph{in-vitro/ex-vivo}\\
Okazawa \emph{et al}. \cite{okazawa2006methods}& 2006& 2D & needle biopsy/drug delivery&\emph{in-vitro}\\
Linguraru \emph{et al}. \cite{linguraru2006texture}& 2006& 3D & tracardiac operation&\emph{in-vitro/in-vivo}\\
Zhou \emph{et al}. \cite{zhou2007automatic}& 2007& 3D & RF-ablation for uterine operation&\emph{in-vitro}\\
Zhou \emph{et al}. \cite{zhou2008automatic}& 2008& 3D & RF-ablation for uterine operation&\emph{in-vitro}\\
Barva \emph{et al}. \cite{barva2008parallel}& 2008& 3D & biopsy/neurological&simulation\\
Aboofazeli \emph{et al}. \cite{aboofazeli2009new}& 2009& 3D & biopsy&\emph{in-vitro}\\
Zhao \emph{et al}. \cite{zhao2009needle}& 2009& 3D & biopsy&\emph{in-vitro}\\
Ren \emph{et al}. \cite{ren2011tubular}& 2011& 3D & catheter intervention&\emph{in-vitro}\\
Zhao \emph{et al}. \cite{zhao2013new}& 2013& 3D & needle biopsy&simulation\\
Cao \emph{et al}. \cite{cao2013automated}& 2013& 3D & catheter for cardiac/prostate/biopsy&\emph{in-vivo}\\
Mohareri \emph{et al}. \cite{mohareri2013automatic}& 2013& 3D+t &prostate therapy&\emph{in-vitro/ex-vivo/in-vivo}\\
Zhao \emph{et al}. \cite{zhao2013automatic}& 2013& 3D+t &needle biopsy&simulation\\
Qiu \emph{et al}. \cite{qiu2013needle}& 2013& 3D &prostate therapy&\emph{in-vitro/ex-vivo/in-vivo}\\
Malekian \emph{et al}. \cite{malekian2014noise}& 2014& 3D &catheter biopsy&simulation\\
Qiu \emph{et al}. \cite{qiu2014phase}& 2014& 3D & prostate therapy &\emph{in-vitro/ex-vivo/in-vivo}\\
Kaya \emph{et al}. \cite{kaya2014gabor}& 2014& 2D &needle biopsy/drug delivery&\emph{in-vitro}\\
Kaya \emph{et al}. \cite{kaya2014needle}& 2014& 2D &needle biopsy/drug delivery&\emph{in-vitro}\\
Beigi \emph{et al}. \cite{beigi2014needle}& 2014& 2D+t &needle biopsy/anesthesia/delivery&\emph{in-vitro/ex-vivo}\\
Beigi \emph{et al}. \cite{beigi2015needle}& 2015& 2D+t &needle biopsy/anesthesia/delivery&\emph{in-vitro/in-vivo}\\
Hacihaliloglu \emph{et al}. \cite{hacihaliloglu2015projection}& 2015& 2D &needle biopsy/anesthesia/therapy&\emph{ex-vivo}\\
Kaya \emph{et al}. \cite{kaya2015real}& 2015& 2D+t &needle biopsy/drug delivery&\emph{in-vitro}\\
Pourtaherian \emph{et al}. \cite{pourtaherian2016automated}& 2016& 3D &needle anesthesia/ablation&\emph{ex-vivo}\\
Beigi \emph{et al}. \cite{beigi2016automatic}& 2016& 2D+t &needle biopsy/nerve block/anesthesias&\emph{in-vitro/in-vivo}\\
Mwikirize \emph{et al}. \cite{mwikirize2016enhancement}& 2016& 2D &needle biopsy/ablation/anesthesia&\emph{ex-vivo}\\
Beigi \emph{et al}. \cite{beigi2016spectral}& 2016& 2D+t &needle biopsy/nerve block/anesthesia&\emph{in-vivo}\\
Kaya \emph{et al}. \cite{kaya2016visual}& 2016& 2D+t &needle biopsy/drug delivery&\emph{in-vitro}\\
Daoud \emph{et al}. \cite{daoud2018hybrid}& 2018& 3D &needle intervention&\emph{ex-vivo}\\
Daoud \emph{et al}. \cite{daoud2018accurate}& 2018& 2D &needle intervention&\emph{ex-vivo}\\
Agarwal \emph{et al}. \cite{agarwal2019real}& 2019& 2D+t &anesthesia/biopsy/brachytherapy&\emph{in-vitro}\\
\hline
\end{tabular}
\end{table*}

\begin{table*}[tb]
\centering
\caption{Overview of papers using learning-based techniques for instrument detection, ordered by year of publication. Clinical applications are mentioned in the literature.}
\label{TB2}
\begin{tabular}{llllll}
\hline
Reference& Year&Format&Application&Dataset\\\hline
Krefting \emph{et al}. \cite{krefting2007segmentation}& 2007& 2D &prostate biopsy&\emph{in-vivo}\\
Uher{\v{c}}{\'\i}k \emph{et al}. \cite{uhervcik2013line}& 2013& 3D &needle biopsy/brachytherapy&simulation/\emph{in-vitro/in-vivo}\\
Geraldes \emph{et al}. \cite{geraldes2014neural}& 2014& 2D+t &needle anesthesia/biopsy/brachytherapy&\emph{in-vitro}\\
Rocha \emph{et al}. \cite{rocha2014flexible}& 2014& 2D&needle anesthesia/biopsy/brachytherapy&\emph{in-vitro}\\
Pourtaherian \emph{et al}. \cite{pourtaherian2015benchmarking}& 2015& 3D&needle biopsy/ablation/anesthesia&\emph{in-vitro/ex-vivo}\\
Hatt \emph{et al}. \cite{hatt2015enhanced}& 2015& 2D&needle biopsy/nerve block/anesthesia&\emph{ex-vivo/in-vivo}\\
Pourtaherian \emph{et al}. \cite{pourtaherian2015multi}& 2015& 3D&needle biopsy/anesthesia&\emph{ex-vivo}\\
Mathiassen \emph{et al}. \cite{mathiassen2016robust}& 2016& 2D+t&needle ablation/biopsy&\emph{ex-vivo}\\
Pourtaherian \emph{et al}. \cite{pourtaherian2017medical}& 2017& 3D&needle/catheter intervention&\emph{in-vitro/ex-vivo/in-vivo}\\
Beigi \emph{et al}. \cite{beigi2017detection}& 2017& 2D+t&needle biopsy/ablation/anesthesia&\emph{in-vivo}\\
Beigi \emph{et al}. \cite{beigi2017casper}& 2017& 2D+t&needle biopsy/ablation/anesthesia&\emph{in-vivo}\\
Mwikirize \emph{et al}. \cite{mwikirize2017local}& 2017& 3D&needle anesthesia&\emph{ex-vivo}\\
Zanjani \emph{et al}. \cite{zanjani2018coherent}& 2018& 3D&needle biopsy/anesthesia&\emph{ex-vivo}\\
Mwikirize \emph{et al}. \cite{mwikirize2018convolution}& 2018& 2D&needle anesthesia/oncology&\emph{ex-vivo}\\
Yang \emph{et al}. \cite{yang2018feature}& 2018& 3D&cardiac catheterization&\emph{in-vitro/ex-vivo}\\
Younes \emph{et al}. \cite{younes2018automatic}& 2018& 3D&prostate brachytherapy&\emph{in-vivo}\\
Pourtaherian \emph{et al}. \cite{pourtaherian2018localization}& 2018& 3D&needle biopsy/ablation/anesthesia&\emph{ex-vivo}\\
Pourtaherian \emph{et al}. \cite{pourtaherian2018robust}& 2018& 3D&needle biopsy/ablation/anesthesia&\emph{ex-vivo}\\
Yang \emph{et al}. \cite{yang2019catheterJMI}& 2019& 3D&cardiac catheterization&\emph{in-vitro/ex-vivo/in-vivo}\\
Yang \emph{et al}. \cite{yang2019catheter}& 2019& 3D&cardiac catheterization&\emph{ex-vivo}\\
Yang \emph{et al}. \cite{yang2019efficient}& 2019& 3D&cardiac catheterization&\emph{ex-vivo}\\
Yang \emph{et al}. \cite{yang2019ISBI}& 2019& 3D&cardiac catheterization&\emph{ex-vivo}\\
Mwikirize \emph{et al}. \cite{mwikirize2019learning}& 2019& 2D+t&needle biopsy/anesthesia&\emph{in-vitro/ex-vivo}\\
Mwikirize \emph{et al}. \cite{mwikirize2019single}& 2019& 2D+t&needle biopsy/anesthesia&\emph{ex-vivo}\\
Arif \emph{et al}. \cite{arif2019automatic}& 2019& 3D&needle biopsy&\emph{in-vitro/in-vivo}\\
Min \emph{et al}. \cite{min2020feasibility}& 2020& 3D&cardiac catheterization&\emph{ex-vivo}\\
Rodgers \emph{et al}. \cite{rodgers2020automatic}& 2020& 2D/3D&interstitial gynecologic brachytherapy&\emph{in-vitro/in-vivo}\\
Zhang \emph{et al}. \cite{zhang2020multiTMI,zhang2020multi}& 2020& 3D&prostate brachytherapy&\emph{in-vivo}\\
Zhang \emph{et al}. \cite{zhang2020multiMP}& 2020& 3D&prostate brachytherapy&\emph{in-vivo}\\
Zhang \emph{et al}. \cite{zhang2020weakly}& 2020& 3D&prostate brachytherapy&\emph{in-vivo}\\
Lee \emph{et al}. \cite{lee2020ultrasound}& 2020& 2D&needle biopsy&\emph{in-vivo}\\
Gillies \emph{et al.} \cite{gillies2020deep}& 2020& 2D&brachytherapy&\emph{ex-vivo/in-vivo}\\
Anders{\'e}n \emph{et al.} \cite{andersen2020deep}& 2020& 3D&brachytherapy/biopsy&\emph{in-vivo}\\
Yang \emph{et al.} \cite{yang2020efficient}& 2020& 3D&anesthesia/catheterization&\emph{ex-vivo}\\
Yang \emph{et al.} \cite{yang2020deepQ}& 2020& 3D&catheterization&\emph{ex-vivo}\\
Yang \emph{et al}. \cite{yang2019MICCAI,yang2021efficient}&2020& 3D&cardiac catheterization&\emph{ex-vivo/in-vivo}\\ \hline
\end{tabular}
\end{table*}

\subsection{Evaluation metrics and performance}
As for evaluation metrics, the used metrics and their definitions are summarized in Table~\ref{TB3}. Specifically, metrics AE to PE are commonly considered to evaluate the performance of the detection. In contrast, the rest metrics are used to evaluate the segmentation performance. The performance of medical instrument detection algorithms are summarized. in Table~\ref{TB4} and Table~\ref{TB5}. To evaluate the detection performance, several validation methods are used: (1) non-learning-based methods are commonly validated on the dataset as described, i.e. validating the method on the whole dataset; (2) learning-based methods are commonly validated on a testing dataset after using a different training dataset for model training. More specifically, cross-validation like leave-one-out cross-validation (LOOCV), k-fold cross-validation (k-CV), or a straightforward dataset split (S), i.e., dividing the dataset into training, validation, and testing subsets, are commonly used in learning-based methods. It is worth mentioning that the spatial resolution of the US images is not summarized because some of the papers did not include this parameter. Some table cells are empty due to unclear descriptions from the considered papers.

\begin{table}[htbp]
\centering
\caption{Summary of Evaluation Metrics For Instrument Detection}
\label{TB3}
\begin{tabular}{lll}
\hline
Name&Abb.&Definitions\\ \hline\hline
Axis Error&AE&\makecell[l]{the average error of each point on the\\ instrument axis}\\ \hline
Detection Error&DE&location-detection error\\ \hline
Diameter Mismatch&DM&\makecell[l]{mismatch between diameters of\\ detection and target}\\ \hline
Detection Rate&-&rate of successfully detect the instrument\\\hline
End-points Error&EE&\makecell[l]{average mismatch of instrument tip\\ and tail}\\ \hline
Execution Time&t&-\\\hline
Failure Rate&-&rate of failed to detect the instrument\\\hline
Mean Square Error&-&average of squares of the detection error\\\hline\hline
Orientation Error&OE&instrument-axis direction mismatch\\ \hline
\makecell[l]{Root-mean-\\square-distance}&RMSD&\makecell[l]{$\frac{\sqrt{\sum_{i=1}^N\sigma_i^2}}{N}$,\\ where $\sigma$ is the distance error}\\\hline
Success Rate&-&rate of successfully detect the instrument\\\hline
Tip Error&TE&instrument-tip mismatch\\ \hline
Tip-to-plane Error&PE&\makecell[l]{point-plane distances between the end\\-points of the ground-truth needle\\ and the detected plane}\\ \hline
Precision&-&true positive/(true positive+false positive)\\\hline
Recall&-&true positive/(true positive+false negtive)\\\hline
Dice Score&DSC&2$\cdot$Precision$\cdot$Recall/(Precision+Recall)\\\hline
\end{tabular}
\end{table}

\begin{table*}[tb]
\centering
\caption{Performances of papers using non-learning-based techniques for instrument detection. Ordered by year to match Table \ref{TB1}. - means not reported in the paper. }
\label{TB4}
\begin{tabular}{llllll}
\hline
Reference& Year&Image Size&Size of Dataset&Key Metrics&Performance\\\hline
Draper \emph{et al}. \cite{draper2000algorithm}&2000&-&33 images&success rate&90$\%$\\
Smith \emph{et al}. \cite{smith2001three}& 2001& - &18$\times$4 images&TE&0.27 mm \\
Ding \emph{et al}. \cite{ding2002automatic}& 2002&$357\times326\times352$&-&OE/t&$1^{\circ}$/1-3 sec. \\
Novotny \emph{et al}. \cite{novotny2003tool} & 2003 &$128\times160\times64$&-&-&- \\
Ding \emph{et al}. \cite{ding2004projection}& 2004&$357\times326\times352$&6 volumes&EE/OE/t&0.7 mm/$1.2^{\circ}$/13FPS\\
Okazawa \emph{et al}. \cite{okazawa2006methods}& 2006&$482\times398$&10 images&AE&0.2-0.8 mm\\
Linguraru \emph{et al}. \cite{linguraru2006texture}& 2006&-&-&-&- \\
Zhou \emph{et al}. \cite{zhou2007automatic}& 2007&$381\times381\times250$&100 trails&OE/EE/t&$1.93^{\circ}$/2.03 mm/0.22 sec.\\
Zhou \emph{et al}. \cite{zhou2008automatic}& 2008&$381\times381\times250$&100 trails&OE/EE/t&$1.58^{\circ}$/1.76 mm/1.76 sec.\\
Barva \emph{et al}. \cite{barva2008parallel}& 2008&$53\times71\times3100$&8 volumes&AE/TE/t&0.301 mm/0.263 mm/1121 sec.\\
Aboofazeli \emph{et al}. \cite{aboofazeli2009new}&2009&$256\times256\times125$&15 volumes&TE/t&2.8 mm/3 sec.\\
Zhao \emph{et al}. \cite{zhao2009needle}& 2009&$50\times50\times50$&6& AE/DE/t& $<2^{\circ}$/$<$2 voxel/$\sim$1.93 sec.\\
Ren \emph{et al}. \cite{ren2011tubular}& 2011&-&6 trails&DM&$<0.4$ mm\\
Zhao \emph{et al}. \cite{zhao2013new}& 2013&-&-&TE/AE/OE improvement&$>92\%/>72\%>71\%$\\
Cao \emph{et al}. \cite{cao2013automated}& 2013&$180\times130\times35$&26 volumes&TE/t&1.11 mm/0.41 sec.\\
Mohareri \emph{et al}. \cite{mohareri2013automatic}& 2013&-&12&target registration error/t&2.68 mm/5 min.\\
Zhao \emph{et al}. \cite{zhao2013automatic}& 2013&-&-&AE/OE improvement&$>60\%>63\%$\\
Qiu \emph{et al}. \cite{qiu2013needle}& 2013&$264\times376\times630$&40 volumes&OE/TE&$0.8^{\circ}$/1 mm\\
Malekian \emph{et al}. \cite{malekian2014noise}& 2014&$53\times71\times160$&28 volumes&failure percent/t&0-70\%/$<$10 sec.\\
Qiu \emph{et al}. \cite{qiu2014phase}& 2014&$130\times210\times270$&25 volumes&EE/detection rate&1.43mm/84\%\\
McSweeney \emph{et al}. \cite{mcsweeney2014estimation}& 2014&-&-&-&-\\
Kaya \emph{et al}. \cite{kaya2014gabor}& 2014&$640\times480$&723 images&detection rate/t&100\%/0.31 sec.\\
Kaya \emph{et al}. \cite{kaya2014needle}& 2014&$640\times480$&164 images&detection rate/t&100\%/0.234 sec.\\
Beigi \emph{et al}. \cite{beigi2014needle}& 2014&-&-&TE&0.16-5.66 mm\\
Beigi \emph{et al}. \cite{beigi2015needle}& 2015&$2.5\times2.5$ cm$^2$&30 images&TE&0.5-0.7 mm\\
Hacihaliloglu \emph{et al}. \cite{hacihaliloglu2015projection}& 2015&$450\times450$&150 images&TE/t&0.49-0.53 mm/0.8 sec.\\
Kaya \emph{et al}. \cite{kaya2015real}& 2015&$640\times480$&112/54/38 images&OE/TE/t&$1.95^{\circ}$/1.22/$<17$ ms.\\
Pourtaherian \emph{et al}. \cite{pourtaherian2016automated}& 2016&-&100 trails&OE/PE/t&$3.44^{\circ}$/0.66 mm/6.25 sec.\\
Beigi \emph{et al}. \cite{beigi2016automatic}& 2016&-&20 trails& OE/TE&$0.93^{\circ}$/1.53 mm \\
Mwikirize \emph{et al}. \cite{mwikirize2016enhancement}& 2016&$370\times370$&100 images&DE/t&0.3 mm/0.6 sec.\\
Beigi \emph{et al}. \cite{beigi2016spectral}& 2016&-&20 sequences&OE&$2.83^{\circ}$\\
Kaya \emph{et al}. \cite{kaya2016visual}& 2016&$640\times480$&7074 images&t&0.10-0.24 sec.\\
Daoud \emph{et al}. \cite{daoud2018hybrid}& 2018&-&450/45 trails&OE/AE&$3.2-4.6^{\circ}$/4.0-4.4 mm\\
Daoud \emph{et al}. \cite{daoud2018accurate}& 2018&-&117 images&OE/AE/TE&$0.2-0.8^{\circ}$/0.2-0.6 mm/0.3-0.6mm\\
Agarwal \emph{et al}. \cite{agarwal2019real}& 2019&-&$\sim$160 frames&TE&0.598 mm\\
\hline
\end{tabular}
\end{table*}

\begin{table*}[tb]
\centering
\caption{Performances of papers using learning-based techniques for instrument detection. Ordered by year to match Table \ref{TB2}. - means not reported in the paper. $\sim$ means average estimation of images. n is a value based on patients}
\label{TB5}
\begin{tabular}{llllll}
\hline
Reference& Year&Image Size&Size of Dataset&Key Metrics&Performance\\\hline
Krefting \emph{et al}. \cite{krefting2007segmentation}& 2007&-&S:1650/-/4950 images&failure rate&6\%\\
Uher{\v{c}}{\'\i}k \emph{et al}. \cite{uhervcik2013line}& 2013&$273\times383\times208$&S:18/-/3 volumes&failure rate/EE/t&0\%/$<$0.5 mm/$\sim$ 300 sec. \\
Geraldes \emph{et al}. \cite{geraldes2014neural}& 2014&-&S:1335/-/422 images&DE&5.68-39.8 mm\\
Rocha \emph{et al}. \cite{rocha2014flexible}& 2014&$101\times101$&S:2272/568/710 images&mean square error& 0.0066\\
Pourtaherian \emph{et al}. \cite{pourtaherian2015benchmarking}& 2015&$\sim148\times169\times159$&LOOCV: 4/4 volumes&DE/OE/t&0.65-0.9 mm/2.2-$3.5^{\circ}$/73-117 sec.\\
Hatt \emph{et al}. \cite{hatt2015enhanced}& 2015&-&LOOCV:577 images&successful rate/DE&99.8\%/0.19 mm\\
Pourtaherian \emph{et al}. \cite{pourtaherian2015multi}& 2015&-&LOOCV:12 volumes &Precision/Recall&0.32/0.75\\
Mathiassen \emph{et al}. \cite{mathiassen2016robust}& 2016&-&S:512/-/1390 images &95th percentile of DE/t&85\% improvement/35.4 FPS\\
Pourtaherian \emph{et al}. \cite{pourtaherian2017medical}& 2017&$\sim180\times190\times206$&LOOCV:9 volumes&EE/OE/t&0.60-0.68 mm/2.2-$3.7^{\circ}$/$>$120 sec.\\
Beigi \emph{et al}. \cite{beigi2017detection}& 2017&-&S: 10/5/5 videos&OE/DE&$2.12^{\circ}$/1.69 mm\\
Beigi \emph{et al}. \cite{beigi2017casper}& 2017&-&S: 18/6/36 videos&success rate/OE/TE&100\%/$1.28^{\circ}$/0.82 mm\\
Mwikirize \emph{et al}. \cite{mwikirize2017local}& 2017&&S:40/-/40 volumes&Precision/Recall/t/TE&0.88/0.98/3.5 sec./0.44 mm\\
Zanjani \emph{et al}. \cite{zanjani2018coherent}& 2018&$\sim184\times249\times203$&S:8/-/2 $\times3$ volumes&DSC improvement&10-20\%\\
Mwikirize \emph{et al}. \cite{mwikirize2018convolution}& 2018&-&10-folds: 2500 images &DSC/OE/TE/t&0.99/$0.82^{\circ}$/0.23 mm/0.58 sec.\\
Yang \emph{et al}. \cite{yang2018feature}& 2018&$\sim152\times163\times110$&LOOCV:20/10/12 volumes&DSC&0.579-0.744\\
Younes \emph{et al}. \cite{younes2018automatic}& 2018&$765\times575\times65$&9 volumes for EM&TE/OE/t&4.2 mm/$6^{\circ}$/1.25 sec. per needle\\
Pourtaherian \emph{et al}. \cite{pourtaherian2018localization}& 2018&$452\times280\times292$&5-CV: 20 volumes&visibility&improved\\
Pourtaherian \emph{et al}. \cite{pourtaherian2018robust}& 2018&$\sim300\times230\times230$&5-CV: 20/20 volumes&DSC/TE&80-84\%/$<$0.7 mm\\
Yang \emph{et al}. \cite{yang2019catheterJMI}& 2019&$\sim152\times163\times110$&LOOCV:10/10/12/8 volumes&DSC/TE/t&0.52-0.83/1.9-3.0 mm/$\sim$450 sec.\\
Yang \emph{et al}. \cite{yang2019catheter}& 2019&$\sim150\times170\times151$&3-CV: 65 volumes&DSC/EE/t&0.54/2.07 mm/10 sec.\\
Yang \emph{et al}. \cite{yang2019efficient}& 2019&$\sim157\times160\times150$&S:62/-/30 volumes&Precision/Recall/t&0.597/0.686/1.1 sec.\\
Yang \emph{et al}. \cite{yang2019ISBI}& 2019&$128\times128\times128$&3-CV: 25 volumes&DSC/EE&0.577/1.8 mm\\
Mwikirize \emph{et al}. \cite{mwikirize2019learning}& 2019&-&S:5000/1000/700 images&TE/t&0.72 mm/0.094 sec.\\
Mwikirize \emph{et al}. \cite{mwikirize2019single}& 2019&$256\times256$&S:7000/-/500 images&TE/t&0.55 mm/67 FPS\\
Arif \emph{et al}. \cite{arif2019automatic}& 2019&$192\times256\times128$&2-CV: 149 volumes&DE/OE/t&1 mm/ $2^{\circ}$/3-5 FPS\\
Min \emph{et al}. \cite{min2020feasibility}& 2020&$376\times92\times88$&2-CV: 8 volumes&DSC/t&0.673/2.2 sec.\\
Rodgers \emph{et al}. \cite{rodgers2020automatic}& 2020&-&S:210/-/52 images&DE/OE&0.27 mm/$0.5^{\circ}$\\
Zhang \emph{et al}. \cite{zhang2020multiTMI,zhang2020multi}& 2020&$1024\times768\times{N}$&S:70/21&detection rate/DE&95\%/1.01 mm\\
Zhang \emph{et al}. \cite{zhang2020multiMP}& 2020&$\sim1024\times768\times{N}$&5-CV: 23 patients&DE/TE&0.290 mm/0.442 mm\\
Zhang \emph{et al}. \cite{zhang2020weakly}& 2020&$\sim1024\times768\times{N}$&5-CV: 10 patients&DE/TE&0.15 mm/0.44 mm\\
Lee \emph{et al}. \cite{lee2020ultrasound}& 2020&$440\times500$&S: 794/-/202 images&DSC/OE&0.567/$13.3^{\circ}$\\
Gillies \emph{et al}. \cite{gillies2020deep}& 2020&$256\times256$&S: 917/325 images&DSC/TE/AE&0.722/4.4mm/$1.4^{\circ}$\\
Anders{\'e}n \emph{et al.} \cite{andersen2020deep}& 2020& $128\times128\times128$&S: 713/389 treatments&RMSD/t&0.55 mm/$<2$ sec.\\
Yang \emph{et al}. \cite{yang2020efficient}& 2020&$160\times160\times160$&S: 64/30 images&AHD/EE/OE&2.4 voxels/2.3mm/$7.3^{\circ}$\\
Yang \emph{et al}. \cite{yang2020deepQ}& 2020&$160\times160\times160$&S: 60/28 images&DSC/t&0.65/1 sec.\\
Yang \emph{et al}. \cite{yang2019MICCAI,yang2021efficient}& 2020&$\sim157\times160\times150$&S:62/30 and 3-folds:18 volumes&DSC/t&0.705/1 sec.\\\hline
\end{tabular}
\end{table*} 

Although there is no commonly used benchmarking dataset for a fair comparison, the detection performance from the literature shows a satisfactory accuracy for clinical usage as all the performances are consistent by considering similar evaluation metrics. However, in terms of real-time efficiency, which is required in clinical applications for operation guidance, it is far from real-time in most papers. Specifically for papers validated on the \emph{in-vivo} datasets, most of them are far from real clinical use. Most of them were validated on limited clinical data, e.g., around ten patients or even less, without reporting the time efficiency (especially for 3D imaging). 

\section{Discussion}\label{Discussions}
Although the technologies described in the literature provide satisfactory results, challenges and limitations still remain. Existing limitations are discussed below.

\noindent{$\bullet$} \emph{Pre- and post-processing}: When considering the commonly used \emph{segmentation modeling} in both non-data-driven and data-driven methods, this modeling is a straightforward method without complex pre- and post-processing compared to state-of-the-art methods in computer vision. A successful segmentation may lead to a better model fitting and detection results, but this approach was only validated on limited datasets. Therefore, it cannot ensure a generalization and robustness for real clinical applications. For example, ultrasound consoles from different vendors with different ultrasound probes have various ultrasound transmit and receive settings leading to different ultrasound image appearances and qualities. The lack of standardization forms a challenge for the desired robustness and generalization of algorithms. A standardization in pre-processing, such as domain adaptation or image normalization, may be required for future improvement. Instead of the popular RANSAC modeling, standardization of model fitting should be applied to improve the efficiency of the post-processing. Meanwhile, this model-fitting method has performance limitations when the instrument has a geometry different from a curved tube. For example, this occurs when instruments with a ball or circular shape are used in cardiology.

\noindent{$\bullet$} \emph{Data collection}: As for data-driven methods and especially deep learning-based methods, algorithms are trained with information from a large amount of annotated data. For clinical applications, this may be difficult to realize due to the limited access to clinical data. Privacy regulations are in place to protect patient data misuse, and patient informed consent must be present to use clinical data and facilitate the training of supervised learning algorithms. For obtaining the reliability of learning, sufficient data should be available. In addition, interpretation of medical data is complex, and only experts can do this, which makes annotated data even more desired, but difficult to collect because of the annotation cost. Therefore, it is a tedious and expensive solution to train a satisfying model for clinical practice, because of the complex network design and the required large data collection. Besides the above challenges, the generation of an annotated dataset with one user may have a bias and intrinsic variability, which would lead to evaluation bias. Therefore, a multi-annotator framework is required to reduce the groundtruth bias, and consider different US machines. Nevertheless, this consideration would lead to higher data collected expense. 

\noindent{$\bullet$} \emph{Algorithm design}: Besides the above challenges for data collection, how to exploit the available datasets, especially for unlabeled images, is important for the deep-learning-based algorithm design. Several solutions could be considered to decrease annotated effort and exploit unlabeled images. (1) A task-specific CNN designs should be employed to decrease overfitting and the total detection time to support real-time application. (2) Domain adaptation may be a solution to address the dataset limitation, which can train a network based on extensive in-vitro/ex-vivo datasets and adapt the network to the in-vivo dataset, which has a limited clinical data size. This approach is cost-effective than directly train from patient data since it requires less annotation by clinical experts. (3) Semi-supervised or weakly supervised learning approach should be considered for the algorithm design, which requires fewer annotation efforts and can easily extend the training dataset for a better and more stable detection algorithm.

\noindent{$\bullet$} \emph{Clinical applications}: Considering the existing literature survey, most technical researchers focus on the most manageable tasks for US-guided interventions, i.e., needle-based anesthesia delivery or biopsy taking, since these datasets are easier to obtain than with prostate brachytherapy and cardiac catheterizations. This trend also reflects that these interventions are more mature and widely accepted by hospitals. Nevertheless, researchers should consider more cooperation with hospitals and industries to develop instrument detection algorithms for different clinical practices and better surgical outcomes. In addition, a more intensively survey should be conducted on both medical doctors and technical researchers, which can lead to a high agreement of the application scenarios with detailed requirement, e.g., make the algorithm more easier to be implemented in the operation procedures with minimized learning curve and design the algorithm pipeline in a way of following the radiologists' preference.

\noindent{$\bullet$} \emph{Detection performance}: As discussed in this paper, the current literature are emphasizing detection accuracy rather than time-efficiency. For US-guided intervention therapies, real-time execution for device detection or tracking is crucial, but the requirement for real-time detection during interventions is under discussion. Specifically, real-time performance is crucial for instrument tracking or guidance based tasks, such as needle tracking. Alternatively, instrument detection might require less time efficiency, such as finding the instrument from a complex 3D imaging modality. Nevertheless, some of the tracking algorithms are based on a frame-by-frame detection algorithm, which places a crucial requirement for the detection efficiency. Moreover, it is unclear for clinical applications how accurate the detection should be for broad acceptance by radiologists, e.g., whether 1-mm detection accuracy or 0.7 DSC segmentation accuracy is sufficient for specific clinical applications. In the future, a more comprehensive study should be performed under cooperation with interventionists, which should validate the importance and different value settings for detection accuracy and detection efficiency.

\noindent{$\bullet$} \emph{Video tracking}: Recently proposed solutions for device segmentation are based on US images, which implies to perform detection algorithms on a frame-by-frame basis for an eventual application. However, in clinical scenarios, real-time US imaging is captured as a video sequence in both 2D and 3D US formats. To better exploit temporal information, 3D+time data should be investigated in future research, which is currently limited by software and hardware implementations. Moreover, the most detection algorithms are executed with more than one second to find the instrument, which is accepted for instrument localization. Nevertheless, this efficiency is far from the requirement of the operation guidance, which asks for a real-time segmentation or detection result for the radiologist during the operation.

\noindent{$\bullet$} \emph{Benchmarking data}: It is important to create benchmarking datasets for different clinical applications, such as biopsy, anesthesia and catheterization, which enable a fair comparison of the time performance, robustness, and accuracy of the different algorithms. This can lead to a higher diversity of dedicated solutions and their acceptance in the community.

\section{Conclusions}\label{Conclusions}
This article has provided a survey of medical instrument detection technologies in US images to guide various types of clinical operations. The segmentation modeling is used in most of the methods in this area, which also follows the development of computer vision in recent years. Moreover, different clinical applications are reviewed, which show that most topics are related to needle-based interventions since it has been widely used in current clinical practice, ranging from anesthesia to biopsy. In addition, catheter-based methods are less considered, because it is guided by X-ray or other image modalities which provide excellent contrast-to-noise ratio, limiting the need for an automatic segmentation tool. The tables and figures presented for this domain show that the choice of device detection technology follows that of the computer vision area so that it can be concluded that image-based methods for interventions have similar limitations as in different non-medical-related areas. For instrument detection, deep learning has become a standard in research because of its increased detection capabilities, but this choice introduces some limiting aspects. For supervised learning, obtaining sufficient data of high quality with expert annotations is serious problem, which is not easily solved. Although techniques for unsupervised learning are emerging, intervention data is still a scarce and significant expansion of available datasets is certainly needed, apart from specific datasets for benchmarking. Furthermore, there is still a trade-off to be made between 2D and 3D processing of the data. Voxel-based processing is powerful for instrument detection, but it hampers real-time clinical applications due to its high complexity. Summarizing, this paper points out the advantages of US-guided interventions, and discusses the weaknesses, such as device detection of these methods and US-based processing to the foreground. This review encourages other researchers to explore US-guided intervention therapy by image processing and paving the way further for this technology's suitable clinical applications.
\section*{Literature Selection}
Conference proceedings was searched for IPCAI, MICCAI, SPIE Medical Imaging, IEEE ISBI, IEEE IUS, IEEE ICIP and IEEE EMBC based on the title and abstract of papers. Moreover, related major journal articles are also searched for IEEE TMI, IEEE TBME, IEEE JBHI, IEEE TUFFC, Medical Physics, MedIA, IJCARS and Ultrasound in Medicine and Biology. With specified searching input, we considered search string as '(Needle OR Catheter OR Instrument) AND (Detection OR Segmentation OR Localization) AND Ultrasound'. We went over all the searched papers by title and abstract to make sure the content is correct. If there were still some misleading, we went to the main content of the paper to make a decision. We also checked references of the papers to confirm the key publications were not missing.
\bibliographystyle{IEEEtran}
\bibliography{draft}

\begin{thebibliography}{100}
\providecommand{\url}[1]{#1}
\csname url@samestyle\endcsname
\providecommand{\newblock}{\relax}
\providecommand{\bibinfo}[2]{#2}
\providecommand{\BIBentrySTDinterwordspacing}{\spaceskip=0pt\relax}
\providecommand{\BIBentryALTinterwordstretchfactor}{4}
\providecommand{\BIBentryALTinterwordspacing}{\spaceskip=\fontdimen2\font plus
\BIBentryALTinterwordstretchfactor\fontdimen3\font minus
  \fontdimen4\font\relax}
\providecommand{\BIBforeignlanguage}[2]{{%
\expandafter\ifx\csname l@#1\endcsname\relax
\typeout{** WARNING: IEEEtran.bst: No hyphenation pattern has been}%
\typeout{** loaded for the language `#1'. Using the pattern for}%
\typeout{** the default language instead.}%
\else
\language=\csname l@#1\endcsname
\fi
#2}}
\providecommand{\BIBdecl}{\relax}
\BIBdecl

\bibitem{douglas2001ultrasound}
B.~R. Douglas, J.~W. Charboneau, and C.~C. Reading, ``Ultrasound-guided
  intervention: expanding horizons,'' \emph{Radiologic Clinics of North
  America}, vol.~39, no.~3, pp. 415--428, 2001.

\bibitem{germano2002advanced}
I.~M. Germano, \emph{Advanced techniques in image-guided brain and spine
  surgery}.\hskip 1em plus 0.5em minus 0.4em\relax Thieme Medical Publishers,
  Incorporated, 2002.

\bibitem{peters2006image}
T.~M. Peters, ``Image-guidance for surgical procedures,'' \emph{Physics in
  Medicine \& Biology}, vol.~51, no.~14, p. R505, 2006.

\bibitem{cleary2010image}
K.~Cleary and T.~M. Peters, ``Image-guided interventions: technology review and
  clinical applications,'' \emph{Annual review of biomedical engineering},
  vol.~12, pp. 119--142, 2010.

\bibitem{scanlan2001invasive}
K.~A. Scanlan, P.~A. Propeck, and F.~T. Lee~Jr, ``Invasive procedures in the
  female pelvis: value of transabdominal, endovaginal, and endorectal us
  guidance,'' \emph{Radiographics}, vol.~21, no.~2, pp. 491--506, 2001.

\bibitem{hatada2000diagnostic}
T.~Hatada, H.~Ishii, S.~Ichii, K.~Okada, Y.~Fujiwara, and T.~Yamamura,
  ``Diagnostic value of ultrasound-guided fine-needle aspiration biopsy,
  core-needle biopsy, and evaluation of combined use in the diagnosis of breast
  lesions,'' \emph{Journal of the American College of Surgeons}, vol. 190,
  no.~3, pp. 299--303, 2000.

\bibitem{coplen1991ability}
D.~E. Coplen, G.~L. Andrile, J.~J. Yuan, and W.~J. Catalona, ``The ability of
  systematic transrectal ultrasound guided biopsy to detect prostate cancer in
  men with the clinical diagnosis of benign prostatic hyperplasia,'' \emph{The
  Journal of urology}, vol. 146, no.~1, pp. 75--77, 1991.

\bibitem{barrington2013ultrasound}
M.~J. Barrington and R.~Kluger, ``Ultrasound guidance reduces the risk of local
  anesthetic systemic toxicity following peripheral nerve blockade,'' 2013.

\bibitem{sheafor1998abdominal}
D.~H. Sheafor, E.~K. Paulson, C.~M. Simmons, D.~M. DeLong, and R.~C. Nelson,
  ``Abdominal percutaneous interventional procedures: comparison of ct and us
  guidance.'' \emph{Radiology}, vol. 207, no.~3, pp. 705--710, 1998.

\bibitem{machi2001ultrasound}
J.~Machi, S.~Uchida, K.~Sumida, W.~M. Limm, S.~A. Hundahl, A.~J. Oishi, N.~L.
  Furumoto, and R.~H. Oishi, ``Ultrasound-guided radiofrequency thermal
  ablation of liver tumors: percutaneous, laparoscopic, and open surgical
  approaches,'' \emph{Journal of Gastrointestinal Surgery}, vol.~5, no.~5, pp.
  477--489, 2001.

\bibitem{oepkes2007successful}
D.~Oepkes, R.~Devlieger, E.~Lopriore, and F.~Klumper, ``Successful
  ultrasound-guided laser treatment of fetal hydrops caused by pulmonary
  sequestration,'' \emph{Ultrasound in Obstetrics and Gynecology: The Official
  Journal of the International Society of Ultrasound in Obstetrics and
  Gynecology}, vol.~29, no.~4, pp. 457--459, 2007.

\bibitem{arashthesis}
A.~Pourtaherian, ``\BIBforeignlanguage{English}{Robust needle detection and
  visualization for 3d ultrasound image-guided interventions},'' Ph.D.
  dissertation, Department of Electrical Engineering, 9 2018, proefschrift.

\bibitem{xia2015plane}
W.~Xia \emph{et~al.}, ``In-plane ultrasonic needle tracking using a fiber-optic
  hydrophone,'' \emph{Medical Physics}, vol.~42, no.~10, pp. 5983--5991, 2015.

\bibitem{krucker2007electromagnetic}
J.~Kr{\"u}cker, S.~Xu, N.~Glossop, A.~Viswanathan, J.~Borgert, H.~Schulz, and
  B.~J. Wood, ``Electromagnetic tracking for thermal ablation and biopsy
  guidance: clinical evaluation of spatial accuracy,'' \emph{Journal of
  Vascular and Interventional Radiology}, vol.~18, no.~9, pp. 1141--1150, 2007.

\bibitem{nadeau2014intensity}
C.~Nadeau \emph{et~al.}, ``Intensity-based visual servoing for instrument and
  tissue tracking in 3d ultrasound volumes,'' \emph{IEEE TASE}, vol.~12, no.~1,
  pp. 367--371, 2014.

\bibitem{draper2000algorithm}
K.~J. Draper, C.~C. Blake, L.~Gowman, D.~B. Downey, and A.~Fenster, ``An
  algorithm for automatic needle localization in ultrasound-guided breast
  biopsies,'' \emph{Medical physics}, vol.~27, no.~8, pp. 1971--1979, 2000.

\bibitem{beigi2020enhancement}
P.~Beigi, S.~E. Salcudean, G.~C. Ng, and R.~Rohling, ``Enhancement of needle
  visualization and localization in ultrasound,'' \emph{International Journal
  of Computer Assisted Radiology and Surgery}, pp. 1--10, 2020.

\bibitem{ding2002automatic}
M.~Ding, H.~N. Cardinal, W.~Guan, and A.~Fenster, ``Automatic needle
  segmentation in 3d ultrasound images,'' in \emph{Medical Imaging 2002:
  Visualization, Image-Guided Procedures, and Display}, vol. 4681.\hskip 1em
  plus 0.5em minus 0.4em\relax International Society for Optics and Photonics,
  2002, pp. 65--76.

\bibitem{ding2004projection}
M.~Ding and A.~Fenster, ``Projection-based needle segmentation in 3d ultrasound
  images,'' \emph{Computer Aided Surgery}, vol.~9, no.~5, pp. 193--201, 2004.

\bibitem{okazawa2006methods}
S.~H. Okazawa, R.~Ebrahimi, J.~Chuang, R.~N. Rohling, and S.~E. Salcudean,
  ``Methods for segmenting curved needles in ultrasound images,'' \emph{MedIA},
  vol.~10, no.~3, pp. 330--342, 2006.

\bibitem{zhou2007automatic}
H.~Zhou, W.~Qiu, M.~Ding, and S.~Zhang, ``Automatic needle segmentation in 3d
  ultrasound images using 3d hough transform,'' in \emph{MIPPR 2007: Medical
  Imaging, Parallel Processing of Images, and Optimization Techniques}, vol.
  6789.\hskip 1em plus 0.5em minus 0.4em\relax International Society for Optics
  and Photonics, 2007, p. 67890R.

\bibitem{zhou2008automatic}
------, ``Automatic needle segmentation in 3d ultrasound images using 3d
  improved hough transform,'' in \emph{Medical Imaging 2008: Visualization,
  Image-Guided Procedures, and Modeling}, vol. 6918.\hskip 1em plus 0.5em minus
  0.4em\relax International Society for Optics and Photonics, 2008, p. 691821.

\bibitem{qiu2013needle}
W.~Qiu, M.~Yuchi, M.~Ding, D.~Tessier, and A.~Fenster, ``Needle segmentation
  using 3d hough transform in 3d trus guided prostate transperineal therapy,''
  \emph{Medical physics}, vol.~40, no.~4, p. 042902, 2013.

\bibitem{barva2008parallel}
M.~Barva, M.~Uhercik, J.-M. Mari, J.~Kybic, J.-R. Duhamel, H.~Liebgott,
  V.~Hlav{\'a}c, and C.~Cachard, ``Parallel integral projection transform for
  straight electrode localization in 3-d ultrasound images,'' \emph{IEEE
  TUFFC}, vol.~55, no.~7, pp. 1559--1569, 2008.

\bibitem{aboofazeli2009new}
M.~Aboofazeli, P.~Abolmaesumi, P.~Mousavi, and G.~Fichtinger, ``A new scheme
  for curved needle segmentation in three-dimensional ultrasound images,'' in
  \emph{2009 IEEE ISBI}.\hskip 1em plus 0.5em minus 0.4em\relax IEEE, 2009, pp.
  1067--1070.

\bibitem{beigi2014needle}
P.~Beigi and R.~Rohling, ``Needle localization using a moving stylet/catheter
  in ultrasound-guided regional anesthesia: a feasibility study,'' in
  \emph{Medical Imaging 2014: Image-Guided Procedures, Robotic Interventions,
  and Modeling}, vol. 9036.\hskip 1em plus 0.5em minus 0.4em\relax
  International Society for Optics and Photonics, 2014, p. 90362Q.

\bibitem{daoud2018accurate}
M.~I. Daoud, A.~Shtaiyat, A.~R. Zayadeen, and R.~Alazrai, ``Accurate needle
  localization using two-dimensional power doppler and b-mode ultrasound image
  analyses: A feasibility study,'' \emph{Sensors}, vol.~18, no.~10, p. 3475,
  2018.

\bibitem{zhao2017evaluation}
Y.~Zhao, Y.~Shen, A.~Bernard, C.~Cachard, and H.~Liebgott, ``Evaluation and
  comparison of current biopsy needle localization and tracking methods using
  3d ultrasound,'' \emph{Ultrasonics}, vol.~73, pp. 206--220, 2017.

\bibitem{smith2001three}
W.~L. Smith, K.~Surry, G.~Mills, D.~B. Downey, and A.~Fenster,
  ``Three-dimensional ultrasound-guided core needle breast biopsy,''
  \emph{Ultrasound in medicine \& biology}, vol.~27, no.~8, pp. 1025--1034,
  2001.

\bibitem{novotny2003tool}
P.~M. Novotny, J.~W. Cannon, and R.~D. Howe, ``Tool localization in 3d
  ultrasound images,'' in \emph{MICCAI}.\hskip 1em plus 0.5em minus 0.4em\relax
  Springer, 2003, pp. 969--970.

\bibitem{linguraru2006texture}
M.~G. Linguraru and R.~D. Howe, ``Texture-based instrument segmentation in 3d
  ultrasound images,'' in \emph{Medical Imaging 2006: Image Processing}, vol.
  6144.\hskip 1em plus 0.5em minus 0.4em\relax International Society for Optics
  and Photonics, 2006, p. 61443J.

\bibitem{zhao2009needle}
S.~Zhao, W.~Qiu, Y.~Ming, and M.~Ding, ``Needle segmentation in 3d ultrasound
  images based on phase grouping,'' in \emph{MIPPR 2009: Medical Imaging,
  Parallel Processing of Images, and Optimization Techniques}, vol. 7497.\hskip
  1em plus 0.5em minus 0.4em\relax International Society for Optics and
  Photonics, 2009, p. 74971L.

\bibitem{qiu2014phase}
W.~Qiu, M.~Yuchi, and M.~Ding, ``Phase grouping-based needle segmentation in
  3-d trans-rectal ultrasound-guided prostate trans-perineal therapy,''
  \emph{Ultrasound in medicine \& biology}, vol.~40, no.~4, pp. 804--816, 2014.

\bibitem{mcsweeney2014estimation}
I.~McSweeney, B.~Murphy, and W.~M. Wright, ``Estimation of needle tip location
  using ultrasound image processing and hypoechoic markers,'' in \emph{2014
  IEEE IUS}.\hskip 1em plus 0.5em minus 0.4em\relax IEEE, 2014, pp. 1876--1879.

\bibitem{frangi1998multiscale}
A.~F. Frangi, W.~J. Niessen, K.~L. Vincken, and M.~A. Viergever, ``Multiscale
  vessel enhancement filtering,'' in \emph{MICCAI}.\hskip 1em plus 0.5em minus
  0.4em\relax Springer, 1998, pp. 130--137.

\bibitem{ren2011tubular}
H.~Ren and P.~E. Dupont, ``Tubular structure enhancement for surgical
  instrument detection in 3d ultrasound,'' in \emph{2011 Annual International
  Conference of the IEEE Engineering in Medicine and Biology Society}.\hskip
  1em plus 0.5em minus 0.4em\relax IEEE, 2011, pp. 7203--7206.

\bibitem{mohareri2013automatic}
O.~Mohareri, M.~Ramezani, T.~K. Adebar, P.~Abolmaesumi, and S.~E. Salcudean,
  ``Automatic localization of the da vinci surgical instrument tips in 3-d
  transrectal ultrasound,'' \emph{IEEE TBME}, vol.~60, no.~9, pp. 2663--2672,
  2013.

\bibitem{zhao2013automatic}
Y.~Zhao, C.~Cachard, and H.~Liebgott, ``Automatic needle detection and tracking
  in 3d ultrasound using an roi-based ransac and kalman method,''
  \emph{Ultrasonic imaging}, vol.~35, no.~4, pp. 283--306, 2013.

\bibitem{malekian2014noise}
L.~Malekian, H.~A. Talebi, and F.~Towhidkhah, ``A noise adaptive method for
  needle localization in 3d ultrasound images,'' in \emph{2014 Iranian
  Conference on Intelligent Systems (ICIS)}.\hskip 1em plus 0.5em minus
  0.4em\relax IEEE, 2014, pp. 1--5.

\bibitem{agarwal2019real}
N.~Agarwal, A.~K. Yadav, A.~Gupta, and M.~F. Orlando, ``Real-time needle tip
  localization in 2d ultrasound images using kalman filter,'' in \emph{2019
  IEEE/ASME International Conference on Advanced Intelligent Mechatronics
  (AIM)}.\hskip 1em plus 0.5em minus 0.4em\relax IEEE, 2019, pp. 1008--1012.

\bibitem{zhao2013new}
Y.~Zhao, C.~Cachard, and H.~Liebgott, ``A new automatically biopsy needle
  tracking method using 3d ultrasound,'' in \emph{2013 IEEE IUS}.\hskip 1em
  plus 0.5em minus 0.4em\relax IEEE, 2013, pp. 844--847.

\bibitem{pourtaherian2016automated}
A.~Pourtaherian, N.~Mihajlovic, S.~Zinger, H.~H. Korsten, P.~H. de~With,
  J.~Huang, and G.~C. Ng, ``Automated in-plane visualization of steep needles
  from 3d ultrasound data volumes,'' in \emph{2016 IEEE IUS}.\hskip 1em plus
  0.5em minus 0.4em\relax IEEE, 2016, pp. 1--4.

\bibitem{cao2013automated}
K.~Cao, D.~Mills, and K.~A. Patwardhan, ``Automated catheter detection in
  volumetric ultrasound,'' in \emph{2013 IEEE ISBI}.\hskip 1em plus 0.5em minus
  0.4em\relax IEEE, 2013, pp. 37--40.

\bibitem{kaya2014gabor}
M.~Kaya and O.~Bebek, ``Gabor filter based localization of needles in
  ultrasound guided robotic interventions,'' in \emph{2014 IEEE IST}.\hskip 1em
  plus 0.5em minus 0.4em\relax IEEE, 2014, pp. 112--117.

\bibitem{kaya2014needle}
------, ``Needle localization using gabor filtering in 2d ultrasound images,''
  in \emph{2014 IEEE ICRA}.\hskip 1em plus 0.5em minus 0.4em\relax IEEE, 2014,
  pp. 4881--4886.

\bibitem{kaya2015real}
M.~Kaya, E.~Senel, A.~Ahmad, O.~Orhan, and O.~Bebek, ``Real-time needle tip
  localization in 2d ultrasound images for robotic biopsies,'' in \emph{2015
  ICAR}.\hskip 1em plus 0.5em minus 0.4em\relax IEEE, 2015, pp. 47--52.

\bibitem{hacihaliloglu2015projection}
I.~Hacihaliloglu, P.~Beigi, G.~Ng, R.~N. Rohling, S.~Salcudean, and
  P.~Abolmaesumi, ``Projection-based phase features for localization of a
  needle tip in 2d curvilinear ultrasound,'' in \emph{MICCAI}.\hskip 1em plus
  0.5em minus 0.4em\relax Springer, 2015, pp. 347--354.

\bibitem{mwikirize2016enhancement}
C.~Mwikirize, J.~L. Nosher, and I.~Hacihaliloglu, ``Enhancement of needle tip
  and shaft from 2d ultrasound using signal transmission maps,'' in
  \emph{MICCAI}.\hskip 1em plus 0.5em minus 0.4em\relax Springer, 2016, pp.
  362--369.

\bibitem{torr2000mlesac}
P.~H. Torr and A.~Zisserman, ``Mlesac: A new robust estimator with application
  to estimating image geometry,'' \emph{Computer vision and image
  understanding}, vol.~78, no.~1, pp. 138--156, 2000.

\bibitem{kaya2016visual}
M.~Kaya, E.~Senel, A.~Ahmad, and O.~Bebek, ``Visual tracking of biopsy needles
  in 2d ultrasound images,'' in \emph{2016 IEEE ICRA}.\hskip 1em plus 0.5em
  minus 0.4em\relax IEEE, 2016, pp. 4386--4391.

\bibitem{beigi2015needle}
P.~Beigi, T.~Salcudean, R.~Rohling, V.~A. Lessoway, and G.~C. Ng, ``Needle
  detection in ultrasound using the spectral properties of the displacement
  field: a feasibility study,'' in \emph{Medical Imaging 2015: Image-Guided
  Procedures, Robotic Interventions, and Modeling}, vol. 9415.\hskip 1em plus
  0.5em minus 0.4em\relax International Society for Optics and Photonics, 2015,
  p. 94150U.

\bibitem{beigi2016automatic}
P.~Beigi, S.~E. Salcudean, R.~Rohling, and G.~C. Ng, ``Automatic detection of a
  hand-held needle in ultrasound via phased-based analysis of the tremor
  motion,'' in \emph{Medical Imaging 2016: Image-Guided Procedures, Robotic
  Interventions, and Modeling}, vol. 9786.\hskip 1em plus 0.5em minus
  0.4em\relax International Society for Optics and Photonics, 2016, p. 97860I.

\bibitem{beigi2016spectral}
P.~Beigi, R.~Rohling, S.~E. Salcudean, and G.~C. Ng, ``Spectral analysis of the
  tremor motion for needle detection in curvilinear ultrasound via
  spatiotemporal linear sampling,'' \emph{IJCARS}, vol.~11, no.~6, pp.
  1183--1192, 2016.

\bibitem{daoud2018hybrid}
M.~I. Daoud, A.-L. Alshalalfah, O.~A. Mohamed, and R.~Alazrai, ``A hybrid
  camera-and ultrasound-based approach for needle localization and tracking
  using a 3d motorized curvilinear ultrasound probe,'' \emph{MedIA}, vol.~50,
  pp. 145--166, 2018.

\bibitem{zanjani2018coherent}
F.~G. Zanjani, A.~Pourtaherian, X.~Tang, S.~Zinger, N.~Mihajlovic, G.~C. Ng,
  H.~H. Korsten \emph{et~al.}, ``Coherent needle detection in ultrasound
  volumes using 3d conditional random fields,'' in \emph{Medical Imaging 2018:
  Image-Guided Procedures, Robotic Interventions, and Modeling}, vol.
  10576.\hskip 1em plus 0.5em minus 0.4em\relax International Society for
  Optics and Photonics, 2018, p. 105760W.

\bibitem{krefting2007segmentation}
D.~Krefting, B.~Haupt, T.~Tolxdorff, C.~Kempkensteffen, and K.~Miller,
  ``Segmentation of prostate biopsy needles in transrectal ultrasound images,''
  in \emph{Medical Imaging 2007: Image Processing}, vol. 6512.\hskip 1em plus
  0.5em minus 0.4em\relax International Society for Optics and Photonics, 2007,
  p. 65122Y.

\bibitem{uhervcik2013line}
M.~Uher{\v{c}}{\'\i}k, J.~Kybic, Y.~Zhao, C.~Cachard, and H.~Liebgott, ``Line
  filtering for surgical tool localization in 3d ultrasound images,''
  \emph{Computers in biology and medicine}, vol.~43, no.~12, pp. 2036--2045,
  2013.

\bibitem{hatt2015enhanced}
C.~R. Hatt, G.~Ng, and V.~Parthasarathy, ``Enhanced needle localization in
  ultrasound using beam steering and learning-based segmentation,''
  \emph{Computerized Medical Imaging and Graphics}, vol.~41, pp. 46--54, 2015.

\bibitem{suykens1999least}
J.~A. Suykens and J.~Vandewalle, ``Least squares support vector machine
  classifiers,'' \emph{Neural processing letters}, vol.~9, no.~3, pp. 293--300,
  1999.

\bibitem{pourtaherian2015benchmarking}
A.~Pourtaherian, S.~Zinger, H.~H. Korsten, N.~Mihajlovic \emph{et~al.},
  ``Benchmarking of state-of-the-art needle detection algorithms in 3d
  ultrasound data volumes,'' in \emph{Medical Imaging 2015: Image-Guided
  Procedures, Robotic Interventions, and Modeling}, vol. 9415.\hskip 1em plus
  0.5em minus 0.4em\relax International Society for Optics and Photonics, 2015,
  p. 94152B.

\bibitem{pourtaherian2015multi}
A.~Pourtaherian, S.~Zinger, N.~Mihajlovic, J.~Huang, G.~C. Ng, H.~H. Korsten
  \emph{et~al.}, ``Multi-resolution gabor wavelet feature extraction for needle
  detection in 3d ultrasound,'' in \emph{Eighth International Conference on
  Machine Vision (ICMV 2015)}, vol. 9875.\hskip 1em plus 0.5em minus
  0.4em\relax International Society for Optics and Photonics, 2015, p. 987513.

\bibitem{pourtaherian2017medical}
A.~Pourtaherian, H.~J. Scholten, L.~Kusters, S.~Zinger, N.~Mihajlovic, A.~F.
  Kolen, F.~Zuo, G.~C. Ng, H.~H. Korsten, and P.~H. de~With, ``Medical
  instrument detection in 3-dimensional ultrasound data volumes,'' \emph{IEEE
  TMI}, vol.~36, no.~8, pp. 1664--1675, 2017.

\bibitem{lafferty2001conditional}
J.~Lafferty, A.~McCallum, and F.~C. Pereira, ``Conditional random fields:
  Probabilistic models for segmenting and labeling sequence data,'' 2001.

\bibitem{mwikirize2017local}
C.~Mwikirize, J.~L. Nosher, and I.~Hacihaliloglu, ``Local phase-based learning
  for needle detection and localization in 3d ultrasound,'' in \emph{Computer
  Assisted and Robotic Endoscopy and Clinical Image-Based Procedures}.\hskip
  1em plus 0.5em minus 0.4em\relax Springer, 2017, pp. 108--115.

\bibitem{younes2018automatic}
H.~Younes, S.~Voros, and J.~Troccaz, ``Automatic needle localization in 3d
  ultrasound images for brachytherapy,'' in \emph{2018 IEEE ISBI}.\hskip 1em
  plus 0.5em minus 0.4em\relax IEEE, 2018, pp. 1203--1207.

\bibitem{yang2018feature}
H.~Yang, A.~Pourtaherian, C.~Shan, A.~F. Kolen \emph{et~al.}, ``Feature study
  on catheter detection in three-dimensional ultrasound,'' in \emph{Medical
  Imaging 2018: Image-Guided Procedures, Robotic Interventions, and Modeling},
  vol. 10576.\hskip 1em plus 0.5em minus 0.4em\relax International Society for
  Optics and Photonics, 2018, p. 105760V.

\bibitem{yang2019catheterJMI}
H.~Yang, C.~Shan, A.~Pourtaherian, A.~F. Kolen \emph{et~al.}, ``Catheter
  segmentation in three-dimensional ultrasound images by feature fusion and
  model fitting,'' \emph{JMI}, vol.~6, no.~1, p. 015001, 2019.

\bibitem{beigi2017detection}
P.~Beigi, R.~Rohling, T.~Salcudean, V.~A. Lessoway, and G.~C. Ng, ``Detection
  of an invisible needle in ultrasound using a probabilistic svm and
  time-domain features,'' \emph{Ultrasonics}, vol.~78, pp. 18--22, 2017.

\bibitem{beigi2017casper}
P.~Beigi, R.~Rohling, S.~E. Salcudean, and G.~C. Ng, ``Casper: computer-aided
  segmentation of imperceptible motion—a learning-based tracking of an
  invisible needle in ultrasound,'' \emph{IJCARS}, vol.~12, no.~11, pp.
  1857--1866, 2017.

\bibitem{mathiassen2016robust}
K.~Mathiassen, D.~Dall’Alba, R.~Muradore, P.~Fiorini, and O.~J. Elle,
  ``Robust real-time needle tracking in 2-d ultrasound images using statistical
  filtering,'' \emph{IEEE Transactions on Control Systems Technology}, vol.~25,
  no.~3, pp. 966--978, 2016.

\bibitem{zhang2020multi}
Y.~Zhang, X.~He, Z.~Tian, J.~Jeong, Y.~Lei, T.~Wang, Q.~Zeng, A.~B. Jani,
  W.~Curran, P.~Patel \emph{et~al.}, ``Multi-needle detection in 3d ultrasound
  images with sparse dictionary learning,'' in \emph{Medical Imaging 2020:
  Ultrasonic Imaging and Tomography}, vol. 11319.\hskip 1em plus 0.5em minus
  0.4em\relax International Society for Optics and Photonics, 2020, p. 113190I.

\bibitem{zhang2020multiTMI}
Y.~Zhang, X.~He, Z.~Tian, J.~J. Jeong, Y.~Lei, T.~Wang, Q.~Zeng, A.~B. Jani,
  W.~J. Curran, P.~Patel \emph{et~al.}, ``Multi-needle detection in 3d
  ultrasound images using unsupervised order-graph regularized sparse
  dictionary learning,'' \emph{IEEE TMI}, 2020.

\bibitem{litjens2017survey}
G.~Litjens, T.~Kooi, B.~E. Bejnordi, A.~A.~A. Setio, F.~Ciompi, M.~Ghafoorian,
  J.~A. Van Der~Laak, B.~Van~Ginneken, and C.~I. S{\'a}nchez, ``A survey on
  deep learning in medical image analysis,'' \emph{MedIA}, vol.~42, pp. 60--88,
  2017.

\bibitem{geraldes2014neural}
A.~A. Geraldes and T.~S. Rocha, ``A neural network approach for flexible needle
  tracking in ultrasound images using kalman filter,'' in \emph{5th IEEE
  RAS/EMBS International Conference on Biomedical Robotics and
  Biomechatronics}.\hskip 1em plus 0.5em minus 0.4em\relax IEEE, 2014, pp.
  70--75.

\bibitem{rocha2014flexible}
T.~S. Rocha and A.~A. Geraldes, ``Flexible needles detection in ultrasound
  images using a multi-layer perceptron network,'' in \emph{5th ISSNIP-IEEE
  Biosignals and Biorobotics Conference (2014): Biosignals and Robotics for
  Better and Safer Living (BRC)}.\hskip 1em plus 0.5em minus 0.4em\relax IEEE,
  2014, pp. 1--5.

\bibitem{pourtaherian2017improving}
A.~Pourtaherian, F.~G. Zanjani, S.~Zinger, N.~Mihajlovic, G.~Ng, H.~Korsten
  \emph{et~al.}, ``Improving needle detection in 3d ultrasound using
  orthogonal-plane convolutional networks,'' in \emph{MICCAI}.\hskip 1em plus
  0.5em minus 0.4em\relax Springer, 2017, pp. 610--618.

\bibitem{pourtaherian2018robust}
A.~Pourtaherian, F.~G. Zanjani, S.~Zinger, N.~Mihajlovic, G.~C. Ng, H.~H.
  Korsten \emph{et~al.}, ``Robust and semantic needle detection in 3d
  ultrasound using orthogonal-plane convolutional neural networks,''
  \emph{IJCARS}, vol.~13, no.~9, pp. 1321--1333, 2018.

\bibitem{yang2018catheter}
H.~Yang, C.~Shan, A.~F. Kolen, and P.~H. de~With, ``Catheter detection in 3d
  ultrasound using triplanar-based convolutional neural networks,'' in
  \emph{2018 25th IEEE ICIP}.\hskip 1em plus 0.5em minus 0.4em\relax IEEE,
  2018, pp. 371--375.

\bibitem{min2020feasibility}
L.~Min, H.~Yang, C.~Shan, A.~F. Kolen \emph{et~al.}, ``Feasibility study of
  catheter segmentation in 3d frustum ultrasounds by dcnn,'' in \emph{Medical
  Imaging 2020: Image-Guided Procedures, Robotic Interventions, and Modeling},
  vol. 11315.\hskip 1em plus 0.5em minus 0.4em\relax International Society for
  Optics and Photonics, 2020, p. 1131521.

\bibitem{yang2019catheter}
H.~Yang, C.~Shan, A.~F. Kolen \emph{et~al.}, ``Catheter localization in 3d
  ultrasound using voxel-of-interest-based convnets for cardiac intervention,''
  \emph{IJCARS}, vol.~14, no.~6, pp. 1069--1077, 2019.

\bibitem{mwikirize2018convolution}
C.~Mwikirize, J.~L. Nosher, and I.~Hacihaliloglu, ``Convolution neural networks
  for real-time needle detection and localization in 2d ultrasound,''
  \emph{IJCARS}, vol.~13, no.~5, pp. 647--657, 2018.

\bibitem{ronneberger2015u}
O.~Ronneberger, P.~Fischer, and T.~Brox, ``U-net: Convolutional networks for
  biomedical image segmentation,'' in \emph{MICCAI}.\hskip 1em plus 0.5em minus
  0.4em\relax Springer, 2015, pp. 234--241.

\bibitem{lee2020ultrasound}
J.~Y. Lee, M.~Islam, J.~R. Woh, T.~M. Washeem, L.~Y.~C. Ngoh, W.~K. Wong, and
  H.~Ren, ``Ultrasound needle segmentation and trajectory prediction using
  excitation network,'' \emph{IJCARS}, vol.~15, no.~3, pp. 437--443, 2020.

\bibitem{rodgers2020automatic}
J.~R. Rodgers, D.~J. Gillies, W.~T. Hrinivich, I.~Gyackov, and A.~Fenster,
  ``Automatic needle localization in intraoperative 3d transvaginal ultrasound
  images for high-dose-rate interstitial gynecologic brachytherapy,'' in
  \emph{Medical Imaging 2020: Image-Guided Procedures, Robotic Interventions,
  and Modeling}, vol. 11315.\hskip 1em plus 0.5em minus 0.4em\relax
  International Society for Optics and Photonics, 2020, p. 113150K.

\bibitem{gillies2020deep}
D.~J. Gillies, J.~R. Rodgers, I.~Gyacskov, P.~Roy, N.~Kakani, D.~W. Cool, and
  A.~Fenster, ``Deep learning segmentation of general interventional tools in
  two-dimensional ultrasound images,'' \emph{Medical physics}, 2020.

\bibitem{pourtaherian2018localization}
A.~Pourtaherian, N.~Mihajlovic, F.~GhazvinianZanjani, S.~Zinger, G.~C. Ng,
  H.~H. Korstcn, and P.~H. De~With, ``Localization of partially visible needles
  in 3d ultrasound using dilated cnns,'' in \emph{2018 IEEE IUS}.\hskip 1em
  plus 0.5em minus 0.4em\relax IEEE, 2018, pp. 1--4.

\bibitem{yang2019efficient}
H.~Yang, C.~Shan, A.~F. Kolen, and P.~H. de~With, ``Efficient catheter
  segmentation in 3d cardiac ultrasound using slice-based fcn with deep
  supervision and f-score loss,'' in \emph{2019 IEEE ICIP}.\hskip 1em plus
  0.5em minus 0.4em\relax IEEE, 2019, pp. 260--264.

\bibitem{yang2019ISBI}
H.~Yang, C.~Shan, A.~F. Kolen, and H.~de~With~Peter, ``Improving catheter
  segmentation \& localization in 3d cardiac ultrasound using direction-fused
  fcn,'' in \emph{2019 IEEE ISBI}.\hskip 1em plus 0.5em minus 0.4em\relax IEEE,
  2019, pp. 1122--1126.

\bibitem{yang2019automated}
H.~Yang, C.~Shan, A.~F. Kolen, and P.~H. de~With, ``Automated catheter
  localization in volumetric ultrasound using 3d patch-wise u-net with focal
  loss,'' in \emph{2019 IEEE ICIP}.\hskip 1em plus 0.5em minus 0.4em\relax
  IEEE, 2019, pp. 1346--1350.

\bibitem{yang2019MICCAI}
H.~Yang, C.~Shan, T.~Tan, A.~F. Kolen \emph{et~al.}, ``Transferring from
  ex-vivo to in-vivo: Instrument localization in 3d cardiac ultrasound using
  pyramid-unet with hybrid loss,'' in \emph{MICCAI}.\hskip 1em plus 0.5em minus
  0.4em\relax Springer, 2019, pp. 263--271.

\bibitem{yang2020deepQ}
H.~Yang, C.~Shan, A.~F. Kolen, and P.~H. de~With, ``Deep q-network-driven
  catheter segmentation in 3d us by hybrid constrained semi-supervised learning
  and dual-unet,'' \emph{arXiv preprint arXiv:2006.14702}, 2020.

\bibitem{yang2021efficient}
H.~Yang, C.~Shan, A.~Bouwman, A.~F. Kolen, and P.~H. de~With, ``Efficient and
  robust instrument segmentation in 3d ultrasound using
  patch-of-interest-fusenet with hybrid loss,'' \emph{Medical Image Analysis},
  vol.~67, p. 101842, 2020.

\bibitem{zhang2020multiMP}
Y.~Zhang, Y.~Lei, R.~L. Qiu, T.~Wang, H.~Wang, A.~B. Jani, W.~J. Curran,
  P.~Patel, T.~Liu, and X.~Yang, ``Multi-needle localization with attention
  u-net in us-guided hdr prostate brachytherapy,'' \emph{Medical Physics},
  2020.

\bibitem{andersen2020deep}
C.~Anders{\'e}n, T.~Ryd{\'e}n, P.~Thunberg, and J.~H. Lagerl{\"o}f, ``Deep
  learning-based digitization of prostate brachytherapy needles in ultrasound
  images,'' \emph{Medical physics}, 2020.

\bibitem{arif2019automatic}
M.~Arif, A.~Moelker, and T.~van Walsum, ``Automatic needle detection and
  real-time bi-planar needle visualization during 3d ultrasound scanning of the
  liver,'' \emph{MedIA}, vol.~53, pp. 104--110, 2019.

\bibitem{yang2020efficient}
H.~Yang, C.~Shan, A.~Kolen, and P.~H. de~With, ``Efficient medical instrument
  detection in 3d volumetric ultrasound data,'' \emph{IEEE TBME}, 2020.

\bibitem{mwikirize2019single}
C.~Mwikirize, J.~L. Nosher, and I.~Hacihaliloglu, ``Single shot needle tip
  localization in 2d ultrasound,'' in \emph{MICCAI}.\hskip 1em plus 0.5em minus
  0.4em\relax Springer, 2019, pp. 637--645.

\bibitem{mwikirize2019learning}
------, ``Learning needle tip localization from digital subtraction in 2d
  ultrasound,'' \emph{IJCARS}, vol.~14, no.~6, pp. 1017--1026, 2019.

\bibitem{zhang2020weakly}
Y.~Zhang, J.~Harms, Y.~Lei, T.~Wang, T.~Liu, A.~B. Jani, W.~J. Curran,
  P.~Patel, and X.~Yang, ``Weakly supervised multi-needle detection in 3d
  ultrasound images with bidirectional convolutional sparse coding,'' in
  \emph{Medical Imaging 2020: Ultrasonic Imaging and Tomography}, vol.
  11319.\hskip 1em plus 0.5em minus 0.4em\relax International Society for
  Optics and Photonics, 2020, p. 1131914.

\bibitem{yang2020deep}
H.~Yang, C.~Shan, A.~F. Kolen \emph{et~al.}, ``Deep q-network-driven catheter
  segmentation in 3d us by hybrid constrained semi-supervised learning and
  dual-unet,'' in \emph{International Conference on Medical Image Computing and
  Computer-Assisted Intervention}.\hskip 1em plus 0.5em minus 0.4em\relax
  Springer, 2020, pp. 646--655.

\bibitem{patel2019ultrasound}
S.~A. Patel, K.~Pierko, and R.~Franco-Sadud, ``Ultrasound-guided bedside core
  needle biopsy: A hospitalist procedure team’s experience,'' \emph{Cureus},
  vol.~11, no.~1, 2019.

\bibitem{banerjee2017use}
S.~Banerjee, T.~Kataria, D.~Gupta, S.~Goyal, S.~S. Bisht, T.~Basu, and
  A.~Abhishek, ``Use of ultrasound in image-guided high-dose-rate
  brachytherapy: enumerations and arguments,'' \emph{Journal of contemporary
  brachytherapy}, vol.~9, no.~2, p. 146, 2017.

\bibitem{jensen1996field}
J.~A. Jensen, ``Field: A program for simulating ultrasound systems,'' in
  \emph{10TH NORDICBALTIC CONFERENCE ON BIOMEDICAL IMAGING, VOL. 4, SUPPLEMENT
  1, PART 1: 351--353}.\hskip 1em plus 0.5em minus 0.4em\relax Citeseer, 1996.

\bibitem{treeby2010k}
B.~E. Treeby and B.~T. Cox, ``k-wave: Matlab toolbox for the simulation and
  reconstruction of photoacoustic wave fields,'' \emph{Journal of biomedical
  optics}, vol.~15, no.~2, p. 021314, 2010.

\end{thebibliography}
\end{document}